\documentclass[aps,prd,twocolumn,groupedaddress,showpacs,fleqn]{revtex4}
\usepackage{graphicx}
\usepackage{epsfig}

\usepackage{mathrsfs}
\usepackage{amssymb}
\usepackage[intlimits]{amsmath}

\begin{document}
\numberwithin{equation}{section}
\title{Spatial QCD theory of the dressing of quarks and the origin of three generations}


\author{J. M. Greben}

\email[]{jmgreben@gmail.com}
\altaffiliation{Some of this work was carried out before 2012, while the author was principal scientist at the Council of Scientific and Industrial Research (CSIR) in Pretoria, South Africa}


\begin{abstract}
The dressing of bare massless quarks is described with a spatial theory based on the self-consistent solution of the QCD field equations. After quantization these equations are expressed in terms of quark and gluon creation and annihilation operators and admit a surprisingly elegant exact (operator) solution which eliminates any multi-quark admixtures in the state vector. Hence, this theory is uniquely and exclusively suited to describe the dressing of single bare quarks.
After factorizing out the operators, a finite set of coupled non-linear differential equations results for the reduced c-number quark and gluon fields. These yield three distinct solitonlike solutions, corresponding to the three observed quark generations. Physically each solution represents a quark absolutely confined by the gluon potentials it generates. The radii of the three generations are given by (2.0428..., 1 , 1)/E, while the binding energies are linked to the SU(3) structure constants and given by E - (32/9, 16/9, 1)E, where E sets the energy scale of the system.
To stabilize the system general relativity is required, putting E in the Planck domain. After the introduction of a vacuum term characterized by the cosmological constant the dressed quark mass can be expressed in terms of the gravitational and cosmological constant, nevertheless lying in the MeV range. After the inclusion of the other gauge interactions this theory might well serve as a theoretical laboratory for quantitative tests of the unification of general relativity and QFT in a constrained Planck scale environment.

\end{abstract}
\maketitle
\def\thesection{\arabic{section}}
\section{Introduction}
\label{sec:Introduction}
In the Standard Model (SM) quarks and leptons are considered pointlike elementary particles whose masses are determined experimentally. The need for these and other phenomenological parameters is one of the unsatisfactory aspects of this model. Historically, renormalization provides the theoretical vehicle for the introduction of empirical mass parameters, although it requires the cancellation of large or infinite contributions. Since then the Higgs mechanism, which gives the vector bosons a mass through spontaneous symmetry breaking, has been linked to the quark and lepton masses as well (\cite{Schwartz}, p. 596). However, this has not reduced the need for empirical (mass) parameters, either.
	
The question whether there exists an underlying theory that can predict these mass parameters is therefore as relevant as it was fifty years ago. In the early days of string theory there were hopes that it could provide such predictions. For example, in a basic text on string theory,  Ref.\ \cite{Green}, the authors state that: \emph{a consistent unified theory of gravity and other forces might someday confront experiment through its implications for already measured quantities like the electron mass or the Cabibbo angle} (\cite{Green}, p. 14). However, this hope has gradually evaporated and the focus of string theory now lies elsewhere.

In this paper we show that a spatial QFT dressing theory can explain the existence of the three generations of quarks and may well form a basis for the prediction of the quark and lepton masses in terms of a few fundamental constants of nature. Our description of dressing does not follow the usual route of incorporating dressing interactions in the scattering series, where they display singular behavior. Rather, we use the quantized field equations for the quark and gluon fields to describe dressed quarks as self-contained spherical systems, consisting of a bare quark in interaction with the boson fields it generates itself. In principle these dressing calculations should precede the application of the Standard Model, as they supply information on the properties of the Standard Model fermions and should tell us which self-interaction diagrams must be excluded from the scattering series. This paper only deals with the fundamental task of constructing the self-contained dressed quark system, and does not address questions regarding the integration of this theory in the Standard Model framework. This would be premature anyway until we have a fuller understanding of the dressing theory and its applicability.

The development of a local dressing theory requires various new QFT methodologies in order to deal with the use and quantization of the QFT field equations. It also requires extensions beyond QFT to include certain consequences of general relativity, implying that QFT itself is unable to provide a fully fundamental basis for particle physics, as it requires empirical particle information.

The field equations in QFT are often dismissed as classical equations, a view which is reinforced by the fact in the classical limit $\hbar \rightarrow 0$ the action integral leads to $\delta S=0$, which is precisely the condition that determines the (classical) Euler-Lagrange equations (\cite{Schwartz}, p. 259). However, it seems very strange that the field equations, which play such a central role in all other fields of physics and are essential for the derivation of Noether currents and energy conservation, would play such a limited role in QFT. Dirac observed as much, when in 1979 he stated that \emph{Methods based on the equations of motion (i.e. the field equations), so necessary for low energy physics have been largely abandoned as being intractable in QFT. Yet if we believe in the unity of physics, we should believe that the same basic ideas universally apply to all fields of physics} \cite {Dirac_Eq_Motion}. As we will see, quantization restores the field equations to their rightful place in QFT and also reveals their special - yet fundamental - role in QFT.

The currently popular Feynman path integral (FPI) quantization method seems not suitable here. The QFT field equations are expressed in terms of a single space-time variable, in which case the notion of paths -- which is so essential for FPI and other scattering approaches -- does not enter in a natural way. The reason for this limitation to a single space-time variable in the Lagrangian (and thus in the field equations) lies in the stringent demands of locality and relativity in QFT.
This restriction limits the applicability of the field equations, as they cannot deal with multi-particle problems in QFT, like they can in non-relativistic quantum mechanics (NRQM). Hence, QFT scattering methods must also be used for multi-particle bound-state problems, implying that they have a much larger domain of validity in QFT than they do in NRQM. An example is the case of lattice gauge calculations, where one deals with multi-particle bound-state problems, but which are based on FPI (\cite{Schwartz}, p. 252). However, this does not mean that the FPI is universally applicable. The dressing of a single bare elementary particle is a bound-state problem that still can -- and also has to -- be dealt with through the quantized field equations. This conclusion is rigorously confirmed by the operator formulation of the field equations, which yield an exact quantum solution which excludes any multi-particle components in the state vector, so that this approach is uniquely equipped to deal with the dressing problem in QFT.

The natural tool for carrying out quantization for this problem is canonical quantization, as it emphasizes the particle - rather than the wave - aspects. By adapting this method to the (discrete) bound-state case we are led to the Wigner-Jordan \mbox{(anti-)commutation} rules for the creation and annihilation operators, which refer to the (initially unknown) quark bound-state wave functions. After the quark fields are thus expanded, consistency demands that the gluon fields are also expanded in these operators, now multiplied with unknown gluon amplitudes. The non-linear field equations then generate higher-order operator terms, so that the fields ultimately turn into infinite sums of \mbox{(in-)finite} products of creation and annihilation operators, multiplying amplitudes with an ever increasing number of indices. This appears to make the construction of a closed solution an impossible goal. However, certain patterns, which appear to be common between the lowest-order equations, suggest an operator solution with an appealingly simple structure. Upon further examination the correctness of this exact operator solution of binomial form can be confirmed. Its properties can easily be identified, the main one being that it eliminates any multi-particle admixtures in the state vector (this limitation holds separately for particles and anti-particles). Hence, the field equation formulation determines its own domain of applicability, namely the dressed quark system characterized by a single quark state vector.

Using this operator solution one can factorize out the creation and annihilation operators from the field equations, reducing it to a set of coupled non-linear c-number differential equations. The operator solution also ensures connectivity, so that only physically meaningful sequences of terms survive in the reduced field equations. The extensive exploitation of self-consistency constraints and explicit and implicit symmetries enable a series of further simplifications which eventually yield a set of one-parameter solutions of the gluon field equations, which in the limit $\alpha_s\rightarrow 0$ even are of analytic form. Combining this result with the quantized Dirac equations constrains the solutions even further, leading to a discretization of the continuous set of solutions, so that eventually a set of four discrete solutions emerges. One of these can be identified with the original bare quark state and would have been the only (trivial) solution had we started with $\alpha_s= 0$. The other three are structural solutions which meet the criteria for dressed quarks.

This bound-state theory is  better behaved than standard QFT, as it does not lead to infinite diagrams and even allows exact non-perturbative solutions (in the limit $\alpha_s\rightarrow 0$). There is no need for gauge fixing terms in the Lagrangian, in fact maintaining the original form of the Lagrangian (apart from ordering issues) appears essential for obtaining analytic solutions. The only infinite quantities that occur are the potentials (and in some cases their wave functions) which ensure the absolute confinement of the bare quark at a finite radius, thereby confirming the self-contained nature of the dressed state whose internal dynamics is insulated from the outside world.

While certain complications of ordinary QFT are thus absent, there are new problems which need to be addressed. These have to do with the presence of (infinite) products of amplitudes at the same space-time point. In standard QFT such products lead to singularities and infinities, especially in expectation values. These are usually removed by the ad hoc imposition of the normal product on these expressions (see \cite{Bogoliubov}, p. 76 or \cite{Itzykson}, p. 111). The use of this tool has also been continued in string theory (see \cite{Green}, p. 91). However, in the dressing theory, which is characterized by (infinite) products of operators at the same point, this recipe fails and we need a more principled approach. The resolution of this problem was discovered early in the development of the dressing theory. It requires the imposition of the so-called $\mathbb{R}$-product \cite{Ordering} on strings of anti-particle operators which share the same space-time variable. This product plays an essential role in obtaining the correct structure of the operator equations and in the construction of the exact operator solution. It also ensures that unphysical infinite terms are eliminated and physically important high order terms (which would be eliminated under the normal product prescription) are preserved in the resulting coupled differential equations. The physical reason why the $\mathbb{R}$-product must be imposed in QFT is that it eliminates the inherent bias towards particles in the calculation of expectation values. It also plays an important role in other problems. For example, it resolves the cosmological constant problem \cite{CC problem}.

Another problem which needs to be addressed is the scale invariance of the field equations. As mentioned earlier the physical scale can only be set after certain effects of general relativity are taken into account. Finally, the negativity of the QFT energy (unacceptable in a scattering formulation, but here acceptable as it represents the binding energy of the bound state) must be countered by a positive vacuum energy to ensure the positivity of the dressed mass. This vacuum energy is supplied by a miniature vacuum universe, again contained within the absolutely confined system.

The outline of the paper is as follows. In the next section we present a short historical perspective on this work and its origins. In Sec.\ \ref{sec:quantization} we discuss the quantization of the field equations, followed by the discussion of the gluon differential equations in Sec.\ \ref{sec:differential equations} and the Dirac equations in Sec.\ \ref{sec:Dirac equations}. In Sec.\ \ref{sec:wave functions} we present the radial behavior of the wave functions and potentials for the three dressing solutions. Then in Sec.\ \ref{sec:TotalEnergy} we discuss the determination of the QFT bound state energy of each solution, a calculation which can be carried out quite independently from the prior dynamical calculations. In Sec.\ \ref{sec:Mass} we indicate which fundamental extensions are needed to make quark mass predictions: taking account of general relativity (GR) and including a (cosmological) vacuum term. Finally, in Sec.\ \ref{sec:conclusions} we summarize our results and discuss new opportunities to gain further insights in QFT and its unification with GR.

\section{Historical perspective on the development of the dressing theory}
\label{sec:Broad}
In the previous section we presented an introduction to the QFT dressing theory. However, the original goal of this research effort was very different. It might be instructive to present a brief history of this research path in order to understand how its ideas originated and developed, as they lead to quite a novel perspective on QFT, while the late emergence of this theory also deserves an explanation.

In the middle eighties we were involved in the field of bag theories (\cite{Hybrid},\cite{pion}), where nucleons are considered as an assembly of three quarks confined in a bag. These theories were advanced by various groups, notably at MIT (\cite{MIT_Group}, \cite{MIT}, \cite{DeTar}) and Triumf \cite{Thomas}, and had met with considerable success. However, its main feature -- the confining bag -– had a phenomenological character and was not derived from QCD. Our goal was to find such a derivation. Since we expected the non-linear nature of QCD to be essential for the derivation of the confinement potential, thereby inspired by the non-linear sigma model (\cite{non-linear sigma},\cite{sigma}), we started out with these non-linear field equations. To make progress it was necessary to expand the quark field $\psi$ in creation and annihilation operators \cite{CapeTown87}. Also, the restriction to quark states - which was common in bag physics - was inadequate, and the anti-quark degrees of freedom needed to be included as well. Self-consistency of the field equations then determined the operator form of the quantized gluon field.

The next important step was the introduction of the $\mathbb{R}$-product, as this explained the mysterious minus signs which appeared necessary in the field equations, if these were to make sense in the presence of anti-particles. Subsequently it was realized that this $\mathbb{R}$-product is a necessary foundational element in QFT to counter the bias towards particles in expectation values.

As stated earlier, the main breakthrough came when we discovered the exact operator solution of the infinite set of coupled operator equations. However, its main property, that the operator solution restricts the application domain to state vectors consisting of a single-quark, excludes its application to the bag model. We had already found indications that certain many-body terms did not seem to fit the formulation and spoiled emergent patterns in the field equations. We could also have been warned by a remark by one of the authors of the MIT bag model, Ken Johnson, who stated in 1975 that the internal quark structure of hadrons could not be related to particles, since that would restrict the description to a non-relativistic framework \cite{Johnson}. These problems can again be attributed to the unique structure of the QFT field equations which depend on a single space-time variable. Because of this they cannot accommodate the multiple variables typical of non-relativistic bound-state treatments of multi-particle systems.

The discovery of this limitation of the field equation approach caused an important U-turn in our research, as we now had to divert our attention from bag physics to the problem of the dressing of single quarks, thereby confronting the conventional treatment of dressing. The transition to the dressing problem could proceed relatively quickly, as many of the essential ingredients -- such as the quantization procedure and the $\mathbb{R}$-product -- had already been developed. Nonetheless, the road towards the current theory was a long one, requiring many new conceptual steps.

Although the original goal had to be abandoned to be replaced by an even more fundamental goal, our findings still bear relevance in the MIT bag context. First, the effective confining mechanism for the first generation of (light) quarks is identical to the phenomenological one used in the MIT bag (except for the scale of course). We find that the potentials are inversely proportional to the quadratic difference between the large and small component of the quark spinor $f^2-g^2$, so that the potential becomes infinite (absolutely confining) when $|f|=|g|$. But this is the same condition that defines the surface in the MIT bag model. So our methodology demonstrates how such an infinite bag can arise from QCD. The MIT bag model also required a volume background (vacuum) term to enable mass calculations, just like we do in our dressing theory.
This shows that the MIT group was able to postulate the main features of confinement purely on general physics grounds and intuition, although in the bag case it was meant to model the interaction between quarks in a colourless state, rather than the self-interactions. Conversely, it shows that our formal dressing methods lead to a picture of dressed quarks that has as strong physical appeal. Or to paraphrase Dirac's comment a little: although the setting is totally different the emerging picture of dressed quarks again confirms the unity of physics and the universality of basic ideas in physics; in this case the spherical nature of the basic constituents of matter.

From a historical point of view it is also interesting to mention that during the early developments of QFT there were intense discussions about the role of self-interactions, which are well described in the book by Schweber on QED \cite{SS}. In ordinary quantum mechanics such interactions are
excluded from the scattering series, however, in QFT the situation is more complex, as there are a multitude of self-interaction diagrams, some of which play an important role in the explanation of the Lamb shift. This debate was settled in favour of the inclusion of all self-interactions - and thus dressing - in the scattering series, despite the singularities arising from this treatment. However, the criticism on these infinities petered out after the introduction of renormalization. The time was not ripe for the development of a local dressing theory anyway, as the nonlinear gauge theories had not yet been formulated and general relativity was not yet seen by many as a necessary ingredient in particle physics.

We must honour Dirac for still pursuing such goals at a time when most particle theorists were involved in developing QFT scattering theory and renormalization techniques. In 1962 he tried to prove that the muon should be considered as an excited state of the electron by developing a spatial QED model of the electron \cite{Dirac_electron}. This is exactly the philosophy which is supported by the results of the dressing theory. He even went as far as to extend this theory to include the effects of gravity \cite{Dirac_electron_GR}. A further motivation of Dirac was that the resulting finiteness of the electron might resolve the infinity problem in QFT, whose solutions he never accepted, as he considered the renormalization process unnatural \cite{Dirac_Renorm}. Unfortunately these efforts were premature, but they attest to Dirac's far-sighted physical intuition. Remarkably, the link between particle physics units and cosmological parameters, which the dressing theory has uncovered, is also a link Dirac speculated on with his fascination for the recurrence of large numbers and ratios in physics, as expressed in his large number hypothesis \cite{DiracLarge}.

\section{Quantization of the field equations}
\label{sec:quantization}

In this section we discuss the quantization procedure and the construction of the operator solution. For some of the technical details we refer back to our earlier paper \cite{QuarkDressing}, which dealt exclusively with the first generation. We also hope to clarify some issues which are better understood now that we have identified all three dressing solutions.

We start with the QCD Lagrangian (see Eqs.\ (16.1-3) in \cite{Peskin}):
\begin{equation}
\label{eq:Lagrangian}
 \mathscr{L}=\bar{\psi}(\textrm{i}\gamma _\mu
 D^\mu-m)\psi-\frac{1}{4}\textbf{F}^{\mu\nu}\bullet
 \textbf{F}_{\mu\nu},
 \end{equation}
where the covariant derivative is defined as follows:
\begin{equation}
\label{eq:Dmu}
D^\mu =\partial^\mu-\frac{1}{2} \textrm{i}g_s
\boldsymbol{\lambda} \bullet\textbf{A}^{\mu}.
 \end{equation}
The field tensor equals:
 \begin{equation}\label{eq:2}
\textbf{F}^{\mu\nu}=\partial^{\mu}\textbf{A}^{\nu}-\partial^{\nu} \textbf{A}^{\mu}+\frac{g_s}{2}
\left( \textbf{A}^{\mu}\times\textbf{A}^{\nu}-\textbf{A}^{\nu}\times \textbf{A}^{\mu}\right),
 \end{equation}
where we used a symmetrized form in preparation for the quantization process. We use a vector notation for the SU(3)-index, so summations over color indices are implied, where applicable. Contrary to the SM, the quark fields do not carry a generation quantum number, as the dressing theory must explain, rather than postulate, its existence. Also, the electro-weak interactions are not yet included, so there is no reference to charge either. The absence of the generation label reflects the (more) fundamental character of the QCD Lagrangian in the dressing theory. A similar reduction will be necessary for the electro-weak interactions if they are included in the dressing theory.

Since we cannot accept pure mass terms in the fundamental dressing Lagrangian, and want to avoid phenomenological parameters anyway, we would like to set the bare quark mass $m$ equal to zero. However, if we immediately set $m=0$ then the theory lacks the richness to evolve into a proper bound-state dressing theory. Hence, we start by assuming that $m > 0$, and take the limit $m\downarrow 0$ at the end. Physically we can justify this procedure by noting that dressed quarks will have mass, so making this assumption can be seen as a preemptive measure to ensure that the quark spinors are parameterized correctly from the start. A similar limiting procedure $\alpha_s\rightarrow 0$ is later proposed for the strong coupling constant, again avoiding the introduction of a phenomenological or empirical parameter.

In the standard (scattering) formulation of QCD (and QED), additional terms are necessary in the Lagrangian, such as the gauge fixing term $(\partial^{\mu}\textbf{A}_{\mu})^2$, the latter being required to enable the construction of propagators. In the Feynman path integral formulation
so-called Faddeev-Popov ghosts and anti-ghosts are also needed to ensure mathematical consistency (\cite{Schwartz}, p. 509).
The dressing theory is not plagued by such technical complications. Instead, it seems crucial to maintain the basic form of the Lagrangian in order to preserve the symmetries that enable the reduction process which eventually leads to the exact solutions.

The classical field equations for the gluon field (Eq.\ (15.51), \cite{Peskin}) read as follows after symmetrization:
\begin{equation}
\label{eq:gluon_field}
\partial_\mu \textbf{F}^{\nu\mu}=\frac{g_s }{2} \bar{\psi}\boldsymbol{\lambda}\gamma^\nu \psi
 +\frac{g_s }{2}\left(\textbf{F}^{\nu\mu} \times \textbf{A}_{\mu}-\textbf{A}_{\mu}\times \textbf{F}^{\nu\mu} \right),
 \end{equation}
while the quark spinor field satisfies the Dirac equation:
\begin{equation}
\label{eq:Dirac}
\left( \textrm{i}\gamma _\mu \partial ^\mu -m \right )\psi(x)=
- \frac{\textrm{i}}{2} g_s\gamma_\mu \lambda_a\psi(x) A_a^\mu(x).
\end{equation}
Our next step is to quantize the fields and field equations. In analogy to the scattering case we start with a linear expansion of the quark field in a (now discrete rather than continuous) set of creation and annihilation operators for the quarks ($b_\alpha^\dag,b_\alpha$) and anti-quarks ($d_\alpha^\dag,d_\alpha$):
\begin{equation}
\label{eq:psi_expansion}
\psi(x)=\sum _{\alpha}  b_\alpha \phi_\alpha(x)+\sum _{\alpha} d_\alpha^\dag \phi_\alpha^a (x),
\end{equation}
where the operators satisfy the following discrete fermionic anti-commutation rules:
\begin{equation}
\label{eq:anti_commutator}
\left\{ b_\alpha,b_\beta^\dag \right\}=\delta_{\alpha,\beta};\left\{ d_\alpha,d_\beta^\dag\right\}=\delta_{\alpha,\beta},
\end{equation}
with all other anti-commutators zero. In scattering theory the time dependence is sometimes absorbed in the creation and annihilation operators, which may be appropriate there, as the scattering states evolve with time. However, in the bound-state case the states are stationary and are characterized by a common stationary time dependence, which can best be carried by the expansion coefficients.

The expansion coefficients $\phi_\alpha(x)$ and $\phi_\alpha^a (x)$ represent normalizable Dirac spinors for quarks and anti-quarks. The wave functions are not known beforehand, in fact at this stage we do not even know whether the dressing equations support bound-state solution(s). Since QCD initiates transitions between colour states and between spin states, the set $\{\alpha\}$ must consist of a complete set of colour $\{\xi_\alpha\}$ and spin states $\{\chi_\alpha\}$. After the introduction of electroweak interactions this set would have to be expanded, but it remains discrete and finite, which is a considerable simplification over the scattering case. External continuous quantum numbers, such as the overall momentum of the dressed system, do not feature here, as we consider the system in its own center-of-mass/energy.

The S-wave Dirac bound-state wave functions are defined in terms of the large and small components $f(r)$ and $g(r)$:
\begin{equation}
\label{eq:5}
\phi_\alpha(\textbf{r},t)=\exp(-\textrm{i}Et)\frac{1}{r \sqrt{4\pi}}
\left( \begin{array}{c}
f(r)\\
-\textrm{i}\boldsymbol{\sigma}\bullet \hat{\textbf{r}}g(r)\\
\end{array} \right)
\chi_\alpha\xi_\alpha,
 \end{equation}
where $r$ is the radial distance to the center-of-energy of the dressed system. The radial wave functions satisfy the normalization condition:
\begin{equation}
\label{eq:norm}
\int^{r_0}_0 dr \ [f^2(r)+g^2(r)]=1,
\end{equation}
where we assumed that the quark is (absolutely) confined within a sphere of radius $r_0$.
The stationary nature of the bound state is characterised by a common time dependence, characterized by the value of $E$ in $\exp(-iEt)$. Since QCD has no dimensionful parameters, the dressing theory is scale invariant and the scale parameter $E$ is as yet undetermined (the mass $m$ does not provide a scale as we set $m\downarrow0$ at the end). In Section \ref{sec:Mass} we will show how this scale can be fixed.

The first quantization assumption, Eq.\ (\ref{eq:psi_expansion}), must now be followed by a sequence of additional quantization steps until full self-consistency is reached. The next step is to insert expansion (\ref{eq:psi_expansion}) in the quark source term in the gluon field equation (\ref{eq:gluon_field}). Ignoring the non-linear gluon terms for now, we find that the gluon field must be expressed in bilinear products of quark operators:
\begin{eqnarray}
\label{eq:bilinear}
\begin{aligned}
& A_a^\mu(x)=\sum_{\alpha,\beta}b_\alpha^\dag b_\beta
A_{a,\alpha\beta}^{\mu,pp}(x)+ \sum_{\alpha,\beta}d_\alpha d_\beta^\dag
A_{a,\alpha\beta}^{\mu,aa}(x)
\\
& +\sum_{\alpha,\beta}b_\alpha^\dag d_\beta^\dag
A_{a,\alpha\beta}^{\mu,pa}(x)+ \sum_{\alpha,\beta}d_\alpha b_\beta
A_{a,\alpha\beta}^{\mu,ap}(x),
\end{aligned}
\end{eqnarray}
for a total of $2^{\text{2}} =4$ distinct operator terms. We thus see that the gluon fields must also be expressed in terms of quark creation and annihilation operators. This is a necessary consequence of the self-consistent use of the full set of field equations and gives the dressing calculation a very different character than it has in scattering theory, where these fields are quantized independently. This has considerable consequences for the ordering of operators, as the gluon and quark fields no longer commute in general.

If we now insert this expansion back into the Dirac equation then the Dirac field acquires terms which are cubic in the quark operators. Inserting these in the gluon field equations and taking account of the non-linear gluon terms, we find that the the gluon fields acquire higher-order terms, which in next order are quartic in the creation and annihilation operators:
\begin{eqnarray}
\label{eq:quartic}
\begin{aligned}
& A_a^\mu(x)\rightarrow
\sum_{\alpha,\beta,\gamma,\delta}b_\alpha^\dag b_\beta b_\gamma^\dag b_\delta
A_{a,\alpha\beta\gamma\delta}^{\mu,pppp}(x)
+\cdots+
\\
& +\sum_{\alpha,\beta,\gamma,\delta}d_\alpha d_\beta^\dag b_\gamma^\dag b_\delta
A_{a,\alpha\beta\gamma\delta}^{\mu,aapp}(x)+\cdots
\end{aligned}
\end{eqnarray}
In this case there is a total of $2^{\text{4}} =16$ distinct operator terms. Hence, the creation and annihilation operators take on a much more dynamic role than they do in scattering theory, where they often are used to  multiply known (classical) solutions of the linear(ized) field equations. With every further iteration an increasing number of quark ($2^{{\text{n+1}}}$) and gluon ($2^{\text{n}}$) profile functions need to be introduced, each multiplying operators of a correspondingly higher order. All these profile functions need to be fixed by additional differential equations and constraints. This process continues indefinitely and suggests that the quantized field equations do not lead to a feasible solution scheme. This might well be the reason why this route of quantizing the QFT equations of motion was never pursued in the past, or else was abandoned prematurely. However, there is a surprising exact solution which comprises all terms up to infinity. But before we can present this solution we need to discuss a few more technical issues.

Since the gluon field and quark field operator components do not commute in general, one has to specify the order of these fields. This can be accomplished by symmetrizing the original classical expressions, while respecting the natural order of the fields, so that $\psi A^{\mu}_a$ and $A^{\mu}_a \bar{\psi}$ are allowed, while ${\psi} A^{\mu}_a$ and $A^{\mu}_a \psi$ are not, as the latter correspond to unphysical sequences of operators. In anticipation of this problem we already wrote the gluon field tensor and the field equations for the gluon and the Dirac field in the correctly symmetrized forms Eqs.\ (\ref{eq:2}), (\ref{eq:gluon_field}) and (\ref{eq:Dirac}). Since the gluon field equations play such a central role in the dressing theory we now present them in fully symmetrized form:
\begin{eqnarray}
\label{eq:F_field_equation}
\begin{aligned}
&-\partial_\mu \partial^\mu \textbf{A}^{\nu} +
\partial^\nu \left(\partial_\mu \textbf{A}^{\mu}\right)
=\frac{g_s }{2} \bar{\psi}\boldsymbol{\lambda} \gamma ^\nu \psi+
\\
&+\frac{g_s}{2}\mathcal{P}_2(A)-\frac{g_s^2}{4}\mathcal{P}_3(A),
\end{aligned}
\end{eqnarray}
where $\mathcal{P}_2(A)$ is quadratic in the gluon fields:
\begin{eqnarray}
\label{eq:P2}
\begin{aligned}
&\mathcal{P}_2(A)=2\left\{\textbf{A}^{\mu} \times (\partial_\mu
\textbf{A}^{\nu}) -(\partial_\mu \textbf{A}^{\nu})\times\textbf{A}^{\mu}\right\}
\\
&+(\partial_\mu \textbf{A}^{\mu})\times
\textbf{A}^{\nu}  -\textbf{A}^{\nu} \times (\partial_\mu \textbf{A}^{\mu}),
\end{aligned}
\end{eqnarray}
and $\mathcal{P}_3(A)$ is cubic in the gluon fields:
\begin{eqnarray}
\label{eq:P3}
\begin{aligned}
\mathcal{P}_3(A)&=
 \textbf{A}_{\mu} \times (\textbf{A}^\nu \times \textbf{A}^{\mu})
-\textbf{A}_{\mu} \times (\textbf{A}^\mu \times \textbf{A}^{\nu})
\\
 -&( \textbf{A}^\nu \times \textbf{A}^{\mu})\times \textbf{A}_{\mu}
 +(\textbf{A}^\mu \times \textbf{A}^{\nu})\times\textbf{A}_{\mu}.
\end{aligned}
\end{eqnarray}
The Dirac equation is already in its correct form, as Eq.\ (\ref{eq:Dirac}) has the right sequence of operators.

Next we must introduce the operator expressions in the field equations and construct the operator solution. The construction of this solution was previously described in Appendix B of Ref.\ \cite{QuarkDressing} and is repeated here because of its central role in demonstrating the feasibility of the dressing theory and because it may have applications beyond the current context.

In order to construct a formal operator solution we can use a simplified form of the field equations, as the operator solution is only affected by the structure of the equations and not by the character of the interaction (except that it must be non-Abelian). So in the analysis of the operator solution we can omit the field indices and the $x$-dependence, so that we can write the field equations schematically as follows:
\begin{eqnarray}
\label{eq:schematic}
H_0\psi=\psi A;\ A=\bar{\psi}\psi+A^2+A^3,
\end{eqnarray}
where we represented the free Dirac Hamiltonian by the symbol $H_0$.
To avoid lengthy expressions, which add little to our understanding, we limit ourselves to particle-terms for the moment. To next-to-lowest-order we then employ the following expansions:
\begin{eqnarray}
\label{eq:2nd order expansion}
\begin{aligned}
&\psi=\sum _{\alpha}b_\alpha \phi_\alpha+\sum _{\alpha\beta\gamma}b_{\alpha}^{\dag} b_{\beta} b_{\gamma}\phi_{\alpha\beta\gamma};
\phi_{\alpha\beta\gamma}=-\phi_{\alpha\gamma\beta},
\\
&A= \sum _{\alpha\beta}b_{\alpha}^{\dag} b_{\beta} A_{\alpha\beta}
+\sum _{\alpha\beta\gamma\delta} b_{\alpha}^{\dag} b_{\beta}^{\dag} b_{\gamma}b_{\delta} A_{\alpha\beta\gamma\delta},
\\
&A_{\alpha\beta\gamma\delta}=-A_{\beta\alpha\gamma\delta}=-A_{\alpha\beta\delta\gamma}=A_{\beta\alpha\delta\gamma}.
\end{aligned}
\end{eqnarray}
Inserting these expansions in Eq.\ (\ref{eq:schematic}) we obtain after a considerable amount of anti-commutator algebra:
\begin{eqnarray}
\begin{aligned}
\label{eq:Ho_1st_order}
H_0\phi_{\alpha}&=\phi_{\beta}A_{\beta\alpha}
\\
H_0\phi_{\alpha\beta\gamma}&=-\frac{1}{2}\phi_{\beta}A_{\alpha\gamma}+\frac{1}{2}\phi_{\gamma}A_{\alpha\beta}
+2\phi_{\epsilon}A_{\epsilon\alpha\beta\gamma}
\\
+&\phi_{\alpha\beta\epsilon}A_{\epsilon\gamma}-\phi_{\alpha\gamma\epsilon}A_{\epsilon\beta}
+2\phi_{\alpha\epsilon\tau}A_{\tau\epsilon\beta\gamma}.
\end{aligned}
\end{eqnarray}
Carrying out such lengthy algebraic manipulations using anti-commutator algebra was quite a common activity in QFT before the FPI method became popular. In Ref.\ \cite{Sakurai} we can find many examples. Also in non-relativistic physics these methods have been applied extensively \cite{Broglia}, also by the author \cite{Greben_PV_model}. Clearly, the number of required manipulations expand exponentially with every further iteration, and soon become unmanageable, unless one uses specially designed algebraic computer programs. Fortunately, we do not have to rely on such techniques, as the lowest order solutions already suggest a complete solution with a particularly elegant form.

We find that the following second-order profile functions, written as an (anti-)symmetrized combination of the first-order ones, solve the equations to the current order:
\begin{eqnarray}
\label{eq:operator_lowest order}
\begin{aligned}
&\phi_{\alpha\beta\gamma}=\frac{1}{2}\left(\phi_{\beta}\delta_{\alpha\gamma}-\phi_{\gamma}\delta_{\alpha\beta}\right),
\\
A_{\alpha\beta\gamma\delta}&=-\frac{1}{4}\left(\delta_{\beta\gamma} A_{\alpha\delta}+\delta_{\delta\alpha} A_{\beta\gamma}\right.
\left.-\delta_{\alpha\gamma} A_{\beta\delta}-\delta_{\beta\delta} A_{\alpha\gamma}\right).
\end{aligned}
\end{eqnarray}
Combining Eqs.\ (\ref{eq:2nd order expansion}) and (\ref{eq:operator_lowest order}) we can write these expressions in a more elegant form:
\begin{eqnarray}
\label{eq:operator_lowest order_Np}
\begin{aligned}
&\psi=(1-N^p)\sum _{\alpha}  b_{\alpha} \phi_{\alpha},
\\
&A=\sum _{\alpha\beta}  b_{\alpha}^{\dag}(1-N^p)b_{\beta}A_{\alpha\beta},
\end{aligned}
\end{eqnarray}
where the quark number operator is indicated by $N^p$:
\begin{equation}
\label{eq:Np}
 N^p=\sum_{\alpha} b_{\alpha}^\dag b_{\alpha}.
 \end{equation}
Going to higher order already becomes a considerable challenge. However, it is still possible to obtain the next order results and one finds that the current result can be expanded in a natural way by the replacement $(1-N^p)\rightarrow(1-N^p)(1-N^p/2)$. The full operator expression then suggests itself, and is given by:
\begin{eqnarray}
\label{eq:operator_solutions}
\begin{aligned}
&\psi=\Lambda_\infty^p \sum _{\alpha}  b_{\alpha} \phi_{\alpha},
\\
&A=\sum _{\alpha\beta}  b_{\alpha}^{\dag}\Lambda_\infty^p b_{\beta}A_{\alpha\beta}.
\end{aligned}
\end{eqnarray}
Here the infinite operator $\Lambda_\infty^p$ is defined as the infinite limit of the finite operator $\Lambda_n^p$:
\begin{eqnarray}
\label{eq:Lambda_p}
 \Lambda^p_n=\frac{1-N^p}{1}\cdots\frac{n-N^p}{n}\equiv
\left(\begin{array}{c}n-N^p\\n\\
\end{array}\right),
 \end{eqnarray}
where we introduced a concise binomial notation for operators, further demonstrating the elegance and beauty of this solution. We will show below that the validity of this solution can easily be proven rigorously.

For the anti-quarks the same derivation can be used, except that the order of the operators must be reversed, as dictated by the $\mathbb{R}$-product \cite{Ordering}. Since the quark and anti-quark operators (anti-)commute, these cases can be treated independently. In this case the $\Lambda$ operator is given by:
\begin{eqnarray}
\label{eq:Lambda_a}
 \Lambda^a_n=\frac{1-N^a}{1}\cdots\frac{n-N^a}{n}\equiv
\left(\begin{array}{c}n-N^a\\n\\
\end{array}\right),
 \end{eqnarray}
 where the anti-quark number operator has the form:
\begin{equation}
\label{eq:Na}
 N^a=-\sum_{\alpha} d_{\alpha} d_{\alpha}^\dag.
 \end{equation}
In order to understand the anti-particle operator algebra and the particular form of $N^a$ one needs to become familiar with the $\mathbb{R}$-product algebra \cite{Ordering}.

The complete solution can now be obtained by combining these operators into a single operator $\Lambda_\infty$:
\begin{eqnarray}
\label{eq:Lambda_inf}
 \Lambda_\infty= \lim_{n\rightarrow\infty}\Lambda^p_n\Lambda^a_n.
\end{eqnarray}
In \cite{QuarkDressing}, the operator $\Lambda_\infty$ was expressed in a terms of a sum of $N^p$ and $N^a$. However, this is only accurate in low order and it should be replaced by the current factorised form.

Returning to the full representation of the amplitudes we can write the exact operator solution of the field equations as follows:
\begin{equation}
\label{eq:quark_expansion_full} \psi(x)=\Lambda_\infty \sum _{\alpha}
\left\{ b_\alpha \phi_\alpha(x)+d_\alpha^\dag \phi_\alpha^a (x)\right\},
\end{equation}
while the gluon field has the form:
\begin{eqnarray}
\label{eq:gluon_expansion_full}
\begin{aligned}
& A_a^\mu(x)=\sum_{\alpha,\beta}\{b_\alpha^\dag \Lambda_\infty b_\beta
A_{a,\alpha\beta}^{\mu,pp}(x)+d_\alpha
\Lambda_\infty d_\beta^\dag A_{a,\alpha\beta}^{\mu,aa}(x)
\\
& +b_\alpha^\dag \Lambda_\infty d_\beta^\dag
A_{a,\alpha\beta}^{\mu,pa}(x)+d_\alpha
\Lambda_\infty b_\beta A_{a,\alpha\beta}^{\mu,ap}(x)\}.
\end{aligned}
\end{eqnarray}
The operator solution is remarkably simple and elegant, especially if one considers that the algebraic derivation becomes unmanageably complex and lengthy after only a few orders. Its validity only depends on the structure of the field equations and is blind to the particular Gauge interaction considered. Hence, it should also apply to the other interactions in the SM. The QFT problem of dressing has now been reduced to the problem of constructing the (finite number of) c-number functions $\phi_\alpha^\ast$ and $A_{a,\alpha\beta}^{\mu,\ast\ast}(x)$, where the superscript $^\ast$ can be either $p$ (particle) or $a$ (anti-particle).

We now discuss a number of important physical properties of this operator solution. One can show that:
 \begin{equation}
\label{eq:identity 1}
 \Lambda^p_\infty \Lambda^p_\infty=\Lambda^p_\infty;\Lambda^a_\infty \Lambda^a_\infty=\Lambda^a_\infty,
 \end{equation}
and
 \begin{equation}
\label{eq:identity 2}
 \Lambda^p_\infty \Lambda^a_\infty=\Lambda^a_\infty \Lambda^p_\infty\neq0.
 \end{equation}
 The identities in Eq.\ (\ref{eq:identity 1}) suggest that the $\Lambda^\ast_\infty$ are projection operators, however, Eq.\ (\ref{eq:identity 2}) seems to contradict this impression. In fact, we now show that these operators project out one-body states, which makes them rather unusual projection operators. To prove this assertion we first note the following properties:
 \begin{equation}
\label{eq:identity 3}
 b_\alpha\Lambda^p_\infty =0; \Lambda^p_\infty b_\alpha^\dag =0;
  d_\alpha^\dag\Lambda^a_\infty =0; \Lambda^a_\infty d_\alpha =0;\forall \alpha.
 \end{equation}
Using these identities it is easy to show that Eqs.\ (\ref{eq:quark_expansion_full}) and (\ref{eq:gluon_expansion_full}) solve the quantized field equations, as we claimed earlier. These identities are also responsible for the connectivity in the resulting coupled differential equations, a very important physical property which gives the resulting differential equations their physical appeal. If the operator $\Lambda_\infty $ is sandwiched between one-particle operators (as it always is in Eq.\ \ref{eq:gluon_expansion_full}), then it acts as a \emph{one-body projection operator} as we already suggested above. Hence, any quark-anti-quark admixtures in the state vector $|b^\dag_\alpha|0\rangle$ are automatically eliminated by the field operators. So after dressing, the quark is still represented by the same single-particle state vector $|b_\alpha^\dag|0\rangle$, except that one now has to specify the additional generation quantum number if there are multiple solutions.

Since the operator solution is determined by the structure of the field equations, this derivation is likely to be applicable to any (non-linear) gauge theory, so the properties derived from this solution (the applicability to single particle state-vectors, connectivity, the number of generations) will likely remain valid once more interactions are introduced.

\section{Reduction of the gluon field equations to differential equations}
\label{sec:differential equations}
After inserting the expansions Eqs.\ (\ref{eq:quark_expansion_full}-\ref{eq:gluon_expansion_full}) in the field equations Eq.\ (\ref{eq:Dirac}) and
(\ref{eq:F_field_equation}), we can eliminate the creation and annihilation operators, thereby creating a more manageable set of c-number equations.

Our next step is to parameterize the gluon amplitudes $A^{\mu,\ast\ast}_{a,\alpha \beta}(x)$, whose structure is determined by the quark source terms in Eq.\ (\ref{eq:F_field_equation}). First we parameterize the color dependence by setting:
\begin{equation}
\label{eq:color_part}
A^{\mu,\ast\ast}_{a,\alpha \beta}(x)=\left(\lambda_a\right)_{\alpha\beta}A^{\mu,\ast\ast}_{\alpha \beta}(x).
\end{equation}
Using the identities:
\begin{equation}
\label{eq:SU3_products}
\sum_{b,c=1}^8f_{abc}\lambda_b\lambda_c=3\textrm{i}\lambda_a;~~
C=\frac{1}{2}\sum_{a=1}^8\lambda_a\lambda_a=\frac{8}{3},
\end{equation}
we can eliminate the color dependence in the field equations. Here we defined the SU(3) Casimir coefficient $C$, which will play an important role in the formalism.

Next the reduced amplitudes $A^{\mu,\ast\ast}_{\alpha \beta}(x)$ are parameterized, again on the basis of the structure of the quark source terms:
\begin{eqnarray}
\label{eq:A0pp}
\begin{aligned}
A^{0,pp}_{\alpha \beta}
&=A^{0,aa}_{a,\alpha\beta}=\frac{F_0(r)}{g_s r}
\delta^{spin}_{\alpha\beta},
\\
\textbf{A}^{pp}_{\alpha \beta}
&=-\textbf{A}^{0,aa}_{\alpha\beta}(\textbf{r})=\frac{F(r)}{g_sr}
(\hat{\textbf{r}}\times \boldsymbol{\sigma})_{\alpha\beta},
\\
A^{0,pa}_{\alpha \beta} &=~~\textrm{i}
\frac{\tilde{F}_0(r)}{g_sr} (\hat{\textbf{r}}\bullet
\boldsymbol{\sigma})_{\alpha\beta}~\exp(2\textrm{i}Et),
\\
\textbf{A}^{pa}_{\alpha \beta}
&=\left[\frac{F_1(r)}{g_sr}\boldsymbol{\sigma}_{\alpha\beta}
+\frac{F_2(r)}{g_sr} \hat{\textbf{r}}(\hat{\textbf{r}}\bullet
\boldsymbol{\sigma})_{\alpha\beta}\right]\exp(2\textrm{i}Et),
\end{aligned}
 \end{eqnarray}
with $A^{0,ap}_{\alpha \beta}$ and $\textbf{A}^{ap}_{\alpha \beta}$ simply Hermitean conjugates of the $pa$ amplitudes.

Inserting these expressions in the gluon field equations allows us to eliminate the dependence on the spin operators as well, resulting in a set of five radial second order differential equations for the gluon fields. This derivation is a straightforward, but tedious exercise, because of all the different spin identities playing a role in the different terms in $\mathcal{P}_2(A)$ and $\mathcal{P}_3(A)$ defined in Eqs.\ (\ref{eq:P2}) and (\ref{eq:P3}).

To simplify the non-linear differential equations for the gluon profile functions we introduce the function $K$, which is the first in a series of auxiliary profile functions, which will help to expose the (hidden) symmetries contained in the symmetrized field equations:
\begin{equation}
\label{eq:K} K=F_1+F_2.
\end{equation}
The five differential equations referring to the profile functions $F_0$, $F$, $\tilde{F}_0$, $F_1$ and $F_2$ now read:
\begin{eqnarray}
\label{eq:F0_equation}
\begin{aligned}
\nonumber
&F_0''=S_0-\frac{6}{r^2}\tilde{F}_0
F_1(1-3F)-\frac{3}{r^2}(2Er+3F_0)\times
\\
&\times(2F_1^2+K^2) -\frac{6}{r^2}K(r\tilde{F}_0'-\tilde{F}_0)
 -\frac{3}{r^2}\tilde{F}_0(rK'+K);
 \end{aligned}
 \end{eqnarray}
\begin{eqnarray}
\label{eq:F_equation}
\begin{aligned}
\nonumber
&F''=S_1-\frac{3}{r^2}\tilde{F}_0F_1(2Er+3F_0)+
\\
&+ \frac{2}{r^2}F(1-\frac{3}{2}F)(1-3F)+\frac{3}{r^2}(\tilde{F}_0^2-F_2^2)(3F-1)
\\
& -\frac{6}{r^2}K(rF_1'-F_1)
+\frac{3}{r^2}F_1(-rK'+K) - \frac{18}{r^2}FF_1K;
\end{aligned}
 \end{eqnarray}
\begin{eqnarray}
\label{eq:crude field equations}
\begin{aligned}
&\tilde{F}_0''=S_1+\frac{2}{r^2}\tilde{F}_0
\left[(1-3F)^2-\frac{9}{2}K^2\right]
\\
& -\frac{6}{r^2}K(rF_0'-F_0)-\frac{1}{r^2}(2Er+3F_0)(rK'+K)+
\\
&+\frac{2}{r^2}F_1(1-3F)(2Er+3F_0);
\\
&F_1''=S_0+\frac{1}{r^2}(1-3F)(rK'-K)-\frac{F_1}{r^2}
\\
&-\frac{F_1}{r^2} (2Er+3F_0)^2-\frac{1}{r^2}\tilde{F}_0(2Er+3F_0)(1-3F)+
\\
&+\frac{1}{r^2}F_1(1-3F)^2-\frac{6}{r^2}F'K-\frac{9}{r^2}F_1K^2-\frac{9}{r^2}F_1^3;
\end{aligned}
 \end{eqnarray}
\begin{eqnarray}
\label{eq:F2_equation}
\begin{aligned}
\nonumber
&\frac{F_2'}{r}-\frac{3F_2}{r^2}=S_2-S_0-\frac{F_2}{r^2}(2Er+3F_0)^2+
\\
&+F_1''-\frac{3F_1'}{r}+\frac{3F_1}{r^2}+\frac{9}{r^2}\tilde{F}_0^2K-(2Er+3F_0)\times
\\
&\times\left(\frac{\tilde{F}_0'}{r}-\frac{\tilde{F}_0}{r^2}\right)+\frac{\tilde{F}_0}{r^2}(2Er+3F_0)(1-3F)+
 \\
&+\frac{3\tilde{F}_0}{r}\left(F_0'-\frac{F_0}{r}\right)+\frac{6}{r}F'F_2-\frac{15}{r^2}FF_2-\frac{9}{r^2}FF_1+
\\
&+\frac{3}{r}F(F_2'+3F_1')+\frac{9}{r^2}F^2(2K-F_1)+\frac{9}{r^2}F_1F_2^2,
\end{aligned}
\end{eqnarray}
where the quark source functions are defined by:
\begin{eqnarray}
\label{eq:25}
\begin{aligned}
S_0&=\frac{\alpha_s}{2}\frac{f^2+g^2}{r},
\\S_1&=\frac{\alpha_s}{2}\frac{2fg}{r},
\\S_2&=\frac{\alpha_s}{2}\frac{f^2-g^2}{r}.
\end{aligned}
\end{eqnarray}
Here the strong coupling constant is given by $\alpha_s=g_s^2/4\pi$.
The source terms are related by the identity:
\begin{equation}
\label{eq:28} S_0^2-S_2^2=S_1^2.
 \end{equation}
The equations contain quadratic and cubic terms with particle and anti-particle intermediate states, with sequences like (pp)(pp) and (pa)(ap). The latter terms are non-classical as they correspond to the creation of quark-anti-quark pairs. They introduce the energy $E$ in the equations. Terms like this, with an intermediate anti-quark state, carry an extra minus sign as a consequence of the $\mathbb{R}$-product. This handy (Feynman-like) prescription captures the main effect of the $\mathbb{R}$-product. Terms like (pa)(pa) or (pp)(aa) are forbidden by the connectivity property of the exact operator solution.

The five coupled equations in (\ref{eq:crude field equations}) do not look very tractable and it seems unlikely that they could be solved, let alone yield an exact analytic solution. However, their structure can be considerably simplified by introducing a new set of auxiliary functions:
\begin{eqnarray}
\label{eq:HG}
\begin{aligned}
F_3&=F_0+\frac{2}{3}Er;~~F_4=F-\frac{1}{3},
\\
H&=3F_3F_1-3F_4\tilde{F}_0;~~G=3F_4^2-3F_1^2-\frac{1}{3}.
\end{aligned}
\end{eqnarray}
After introducing the functional:
\begin{equation}
\label{eq:functional}
Z(A,B,C)=A''+AC^2+2B'C+BC',
\end{equation}
we can cast the first four equations in Eq.\ (\ref{eq:crude field equations}) in the following elegant form:
\begin{eqnarray}
\begin{aligned}
\label{eq:Z1}
Z(F_3,\tilde{F}_0,3K/r)&=S_0-\frac{6}{r^2}F_1H,
\\
 Z(\tilde{F}_0,F_3,3K/r)&=S_1-\frac{6}{r^2}F_4H,
\\
Z(F_4,F_1,3K/r)&=S_1-\frac{3}{r^2}(\tilde{F}_0H-GF_4),
\\
Z(F_1,F_4,3K/r)&=S_0-\frac{3}{r^2}(F_3H-GF_1).
\end{aligned}
\end{eqnarray}
Notice the striking symmetry between $F_3$ and $\tilde{F}_0$ on the one hand, and between $F_1$ and $F_4$ on the other.

A further simplification is possible by adding and subtracting the equations. To this end we introduce another set of auxiliary functions:
\begin{eqnarray}
\label{eq:XYUV}
\begin{aligned}
&X=F_3-\tilde{F}_0; Y=F_3+\tilde{F}_0;
\\
&U=F_4-F_1; V=F_4+F_1,
\end{aligned}
\end{eqnarray}
together with the combined source functions:
\begin{equation}
\label{eq:S+S-}
S_\pm = S_0 \pm S_1;\ S_+S_-=S^2_2.
\end{equation}
The equations in (\ref{eq:Z1}) can then be combined to yield:
\begin{eqnarray}
\begin{aligned}
\label{eq:Z5}
Z(Y,Y,3K/r)&=S_+-\frac{6}{r^2}VH,
\\
Z(X,-X,3K/r)&=S_-+\frac{6}{r^2}UH,
\\
Z(V,V,3K/r)&=S_+-\frac{3}{r^2}(YH-GV),
\\
 Z(U,-U,3K/r)&=-S_-+\frac{3}{r^2}(XH+GU),
\end{aligned}
\end{eqnarray}
 where $G$ and $H$ in Eq.\ (\ref{eq:HG}) can also be written in terms of the new functions:
 \begin{equation}
 \label{eq:Z9}
 H=\frac{3}{2}(VX-UY);G=3UV-\frac{1}{3}.
 \end{equation}
The main reason why these combined equations are more useful than the previous ones is that the $Z$ functional now depends only on two functions and then can be reduced to a pure second order differential:
\begin{eqnarray}
\label{eq:Z_reduction}
Z(A,A,C)=\hat{E}^{-1}\hat{A}'';Z(B,-B,C)=\hat{E}\hat{B}'',
\end{eqnarray}
with $A=\hat{E}^{-1}\hat{A}$, $B=\hat{E}\hat{B}$ and $C=\hat{E}^{-1}\hat{E}'$. To be able to apply these identities we need to set:
\begin{equation}
\label{eq:Ehat}
\hat{E}(r)=\exp \left[{3\int_0^r dr'\frac{K(r')}{r'}}\right]\rightarrow \frac{\hat{E}'}{\hat{E}}=3 \frac{K}{r}.
 \end{equation}
The final equations can now be expressed in terms of hatted functions, which are defined as follows:
\begin{eqnarray}
\begin{aligned}
\label{eq:Xhat}
X=\hat{E}\hat{X};Y&=\hat{E}^{-1}\hat{Y};\ U=\hat{E}\hat{U};V=\hat{E}^{-1}\hat{V},
\\
S_+&=\hat{E}^{-1}\hat{S}_+;\ S_-=\hat{E}\hat{S}_-.
\end{aligned}
\end{eqnarray}
We then find:
\begin{eqnarray}
\begin{aligned}
\label{eq:Z10}
\hat{Y}''&=\hat{S}_+-\frac{6}{r^2}\hat{V}H,
\\
\hat{X}''&=\hat{S}_-+\frac{6}{r^2}\hat{U}H,
 \\
\hat{V}''&=\hat{S}_+-\frac{3}{r^2}(\hat{Y}H-\hat{V}G),
\\
 \hat{U}''&=-\hat{S}_-+\frac{3}{r^2}(\hat{X}H+\hat{U}G).
\end{aligned}
\end{eqnarray}
The functions $H$ and $G$ maintain the same structure as in Eq.\ (\ref{eq:HG}) in the hatted form:
 \begin{equation}
 \label{eq:Z14}
 H=\frac{3}{2}(\hat{V}\hat{X}-\hat{U}\hat{Y});G=3\hat{U}\hat{V}-\frac{1}{3}.
 \end{equation}
The equations in (\ref{eq:Z10}) can easily be solved in the limit $\alpha_s\rightarrow 0$ when the source functions vanish by demanding that $H=G=0$. We write the solution as follows:
\begin{equation}
\label{eq:solutions}
 \hat{U}=\hat{V}=-\frac{1}{3}$  and  $\hat{X}=\hat{Y}=\frac{2}{3}\beta Er.
\end{equation}
We have fixed the value of $\hat{U}$ to $-1/3$, which can be done without lack of generality. The factor $2/3$ in front of the parameter $\beta$ was chosen for future convenience. These solutions will be discussed in more detail in Sec.\ \ref{sec:wave functions}.

Hence, we have dissected the gluon and quark sector by finding solutions for the gluon fields without reference to the quark source functions. Since the gluon fields generate the binding potentials, it looks like we have determined the potentials independent of the wave functions, just like we are used to in non-relativistic physics. However, if we want to solve the Dirac equations we first have to go back to the original gluon profile functions. This requires the knowledge of the bridging function $\hat{E}$, which re-introduces the dependence on the wave functions, as will be discussed in Sec.\ \ref{sec:Dirac equations}.

Before leaving this section we still have to discuss the fifth gluon differential equation in Eq.\ (\ref{eq:crude field equations}) for $F_2$. This equation will help us to determine the unknown function $E$. By adding the last two equations in  Eq.\ (\ref{eq:crude field equations}) one obtains an explicit expression for $K$:
\begin{eqnarray}
\label{eq:K_algebra}
\begin{aligned}
\frac{3K}{r}&=\frac{rS_2/3-F_3\tilde{F}_0'+F_3'\tilde{F}_0-2F_4'F_1+2F_1'F_4}
{F_3^2-\tilde{F}_0^2+2F_1^2-2F_4^2}=
\\
&=\frac{rS_2/3-XY'/2+X'Y/2-U'V+UV'}{XY-2UV}.
\end{aligned}
\end{eqnarray}
It looks as if this equation expresses $K$ in terms of the other four profile functions. However, if we convert this to hatted form $K$ disappears:
\begin{equation}
 \label{eq:K_hatted}
\frac{1}{3}rS_2=\frac{1}{2}(\hat{X}\hat{Y}'-\hat{X}'\hat{Y})+\hat{U}'\hat{V}-\hat{U}\hat{V}'.
 \end{equation}
Hence, this  gluon equation looks more like a consistency condition and $K$ must be determined by other means. Because of the Wronskian nature of the RHS in Eq.\ (\ref{eq:K_hatted}), differentiation of this equation also yields a useful identity. After inserting the second order differential equations from (\ref{eq:Z10}) this identity takes a very simple form which can be expressed in terms of the original functions:
\begin{eqnarray}
\label{eq:rS2}
\begin{aligned}
&\frac{1}{3}(rS_2)'=S_1(F_3+2F_1)-S_0(\tilde{F}_0+2F_4)=
\\
=&S_1(F_0+2F_1)-S_0(\tilde{F}_0+2F)+\frac{2}{3}(ErS_1+S_0).
\end{aligned}
\end{eqnarray}
If one combines this with the Dirac equation it breaks up in two separate equations, as we show in the next section.
\section{Dirac equations for the dressed quarks}
\label{sec:Dirac equations}
After eliminating the operator structure and the color dependence, similar to what was done for the gluon field in Sec.\ \ref{sec:differential equations}, the quantized Dirac equations, Eq.\ (\ref{eq:Dirac}), reduce to two coupled linear c-number differential equations for the large ($f$) and small ($g$) components:
\begin{eqnarray}
\begin{aligned}
\label{eq:fdirac}
\left( E-V-V_s-m\right)f(r)&=-\frac{g(r)}{r}-g'(r)+V_Tg(r),
\\
 \left( E-V+V_s+m\right)g(r)&=-\frac{f(r)}{r}+f'(r)+V_Tf(r).
\end{aligned}
\end{eqnarray}
The potentials are simple linear combinations of the five original gluon profile functions:
\begin{equation}
\label{eq:potentials}
V=-\frac{C}{r}(F_0+2F_1);V_T=\frac{C}{r}(\tilde{F}_0+2F);V_s=-\frac{C}{r}K,
\end{equation}
where $C=\frac{8}{3}$ is the SU(3) constant discussed previously. Three types of potentials are present: vector, scalar and tensor.
One other type of potential is possible in the radial Dirac equations, also being of the vector type \cite{Dirac_potentials}. Such a potential can enter if additional SM interactions are included, such as the Higgs interaction. Its presence would not affect the general formalism in a fundamental way.

If we multiply the equation for $f$ with $g$, and the equation for $g$ with $f$, and add them then we get the following relationship:
\begin{equation}
\label{eq:rS2Dirac}
\frac{1}{3}(rS_2)'-\frac{2}{3}(ErS_1+S_0)=-\frac{2r}{3}(S_1V+S_0V_T).
\end{equation}
If we now insert the expressions for the potentials we get the same expression $S_1(F_0+2F_1)-S_0(\tilde{F}_0+2F)$ as we did in Eq.\ (\ref{eq:rS2}), except that it now features a coefficient $2C/3$, instead of unity. Since $C\neq 3/2$, and both results must be correct, we obtain two separate equations:
\begin{equation}
\label{eq:potential_identity}
S_1V+S_0V_T=0,
\end{equation}
\begin{equation}
\label{eq:cc}
2S_0+2ErS_1=(rS_2)'.
\end{equation}		
The latter equation also follows from current conservation, confirming the correctness of these arguments.  These relationships are an important consequence of the self-consistency between the gluon and quark field equations. For the construction of the final solutions we will make extensive use of Eqs.\ (\ref{eq:potential_identity}) and  (\ref{eq:cc}).

Based on Eq.\ (\ref{eq:potentials}) we can now express the potentials in terms of the unknown function $\hat{E}$ and the known functions $\hat{X}$, $\hat{Y}$, $\hat{U}$, and $\hat{V}$:
\begin{eqnarray}
\begin{aligned}
\label{eq:EV}
V  =&-\frac{C}{r}\left[\frac{\hat{E}\hat{X}+\hat{E}^{-1}\hat{Y}}{2}+\hat{E}^{-1}\hat{V}-\hat{E}\hat{U}-\frac{2}{3}Er\right],
\\
V_s=&-\frac{C}{3}\frac{\hat{E}'}{\hat{E}},
\\
V_T=&\hphantom{-.}\frac{C}{r}\left[\frac{\hat{E}^{-1}\hat{Y}-\hat{E}\hat{X}}{2}+\hat{E}\hat{U}+\hat{E}^{-1}\hat{V}+\frac{2}{3}\right].
\end{aligned}
\end{eqnarray}
The (singular) confining nature of these potentials must reside in the function $\hat{E}$, as the other hatted functions are simple constants or proportional to $r$. In order to determine the potentials and wave functions explicitly we now have to solve for this unknown function $\hat{E}$. This can be done by means of a consistency equation, which also determines the number of allowed physical solutions.

\section{Discussion of the three dressing solutions and their properties}
\label{sec:wave functions}
In order to determine $\hat{E}$ we use a hybrid version of the top equation in Eq.\ (\ref{eq:rS2}), where the original source functions are maintained, but the gluon profile functions are converted to the hatted functions whose solutions are already known:
\begin{eqnarray}
\label{eq:rS2hybrid}
\hat{E}S_+(\frac{\hat{X}}{2}-\hat{U})-\hat{E}^{-1}S_-(\frac{\hat{Y}}{2}+\hat{V})=\frac{1}{3}(rS_2)'.
\end{eqnarray}
Inserting the hatted solutions from Eq.\ (\ref{eq:solutions}) and taking the roots of this quadratic equation in $\hat{E}$, we have:
\begin{eqnarray}
\label{eq:Ehat_root}
\hat{E}=\frac{(rS_2)'\pm\sqrt{\left[(rS_2)'\right]^2+4S_2^2(\beta^2 E^2r^2-1)}}
{2\left(\beta Er+1\right) S_+}.
 \end{eqnarray}
The appearance of a square root in the solution of the field equations is unusual and in our opinion unacceptable, as we do not expect the bound-state solutions to be discontinuous or feature discontinuous derivatives, except of course at the edge of the domain where they might be singular. However, after rewriting the expression inside the square root using Eq.\ (\ref{eq:cc}):
\begin{eqnarray}
\label{eq:Root}
\begin{aligned}
&\left[(rS_2)'\right]^2+4S_2^2(\beta^2 E^2r^2-1) \rightarrow
\\
&4\left[(S_1+ErS_0)^2+(ErS_2)^2(\beta^2-1)\right],
\end{aligned}
\end{eqnarray}
and by setting $\beta^2=1$ we can turn this expression into a pure square, so that the square root can be removed. We then get:
 \begin{eqnarray}
\label{eq:Ehat_regular}
\hat{E}=\frac{S_0+ErS_1+\gamma(S_1+ErS_0)}{S_+(\beta Er+1)};\beta,\gamma=\pm 1,
 \end{eqnarray}
which has the desired analytical behaviour. So instead of a continuum of solutions for arbitrary $\beta$, we get solutions for two discrete values of $\beta$.

It is interesting that this quantization or discretization of the solution space has its origin in the demand of analyticity or continuity; a demand not unfitting these structural/topological solutions and quite different from the usual quantization conditions in non-relativistic physics. This argument was facilitated by the fact that our expressions were exact and mathematically explicit. To impose such a continuity demand in a numerical calculation would be much  more difficult, as discontinuities (especially those in derivatives) are then much harder to spot.

One easily verifies that the inverse of $\hat{E}$ is given by:
\begin{eqnarray}
\label{eq:Ehat_inverse}
\hat{E}^{-1}=\frac{-S_0-ErS_1+\gamma(S_1+ErS_0)}{S_-(\beta Er-1)}.
 \end{eqnarray}
The elegance and simplicity of these equations again underscores the importance of maintaining exactness throughout the solution process.

These equations were derived in the limit $\alpha_s\rightarrow0$. In \cite{QuarkDressing} we showed that $\mathcal{O}(\alpha_s)$ corrections can be added perturbatively, although the combination of non-linear constraints and new integration constants in the perturbative equations offers some challenges. Setting $\alpha_s \neq 0$ requires the introduction of an empirical parameter, which we would like to avoid at this fundamental level. In addition, the fact that the (running) strong coupling constant diminishes in value with increasing energy, may suggest that it reaches a bare value of zero at the Planck level. Presumably this assumption can also be tested during the further development and application of the dressing theory.

Since the only dimensionful parameter in the dressing theory is $E$ (remember $m\rightarrow0$), all equations, formulae and derivatives can be expressed in terms of the dimensionless variable $x$:
\begin{equation}
\label{eq:scalefree}
x=Er.
\end{equation}
Hence, from now on all solutions and associated functions will be presented as functions of $x$. Dimensionful quantities can then simply be multiplied by the appropriate power of $E$. The determination of $E$ cannot be accomplished within the framework of QFT. In Sec.\ \ref{sec:Mass} we show that the inclusion of the effects of general relativity (GR) can fix the value of $E$.

We now get a total of four solutions, each corresponding to one of the pairs ($\beta=\pm 1$, $\gamma=\pm 1$). The first one is:
\vspace{1.5mm}

I. $\beta=1;\gamma=1$.  Bare solution.
\vspace{1.5mm}

\noindent We call this the bare solution as it leads to $\hat{E}=\hat{E}^{-1}=1$, so that the potentials defined in Eq.\ (\ref{eq:EV}) are all zero: $V=V_s=V_T=0$. Hence, this solution does not yield any binding potentials and our original Ansatz, that the solution will represent a bound system, is contradicted. We conclude that the self-interactions of the quark do no necessarily lead to a dressed quark, and that the original bare quark can survive these interactions and maintain its bare status. Since all quarks in the SM are massive, this bare massless quark solution is not part of the SM quark structure. However, since the total number of solutions is $2^2 =4$, this bare solution still plays a role in establishing the number of dressed solution, as this is given by $4-1=3$. This observation may be important if one wants to associate a group structure with the set of three generations.

\vspace{1.5mm}

II. $\beta=-1;\gamma=-1$.  First generation.
\vspace{1.5mm}

\noindent As we will see this solution has the largest binding energy in units of $E$, so we identify this as the first generation of quarks. The function $\hat{E}$ is now given by:
\begin{equation}
\label{eq:Ehat1}
\hat{E}= \frac{S_0+xS_1-S_1-x S_0}{(-x+1)S_+}=\frac{S_-}{S_+}=\frac{(f-g)^2}{(f+g)^2}.
\end{equation}						
This solution was discussed at length in Ref.\ \cite{QuarkDressing}, see Eq.\ (102). From  Eq.\ (\ref{eq:EV}) we then find that the potentials are given by:
\begin{eqnarray}
\begin{aligned}
\label{eq:Vs1}
\frac{V(x)}{E}&=\hphantom{-}\frac{4C}{3x}\frac{f^2+g^2}{(f^2-g^2)^2}\left[x(f^2+g^2)+2fg\right],
\\
\frac{V_s(x)}{E}&=-\frac{4C}{3}\frac{f'g-fg'}{f^2-g^2},
\\
\frac{V_T(x)}{E}&=-\frac{4C}{3x}\frac{2fg}{(f^2-g^2)^2}\left[x(f^2+g^2)+2fg\right].
\end{aligned}
\end{eqnarray}
These potentials become infinite for $x=x_0$, where $x_0$ is defined as the first non-zero point where $f(x_0)=-g(x_0)$. Hence, these potentials confine the (bare) quark absolutely at a radius $x=x_0$ from the centre of the system.

The dependency of the potentials on the wave functions (though not on their normalization) is a beautiful demonstration of the far-reaching consequences of self-consistency in the QFT field equations. Superficially, this makes their solution even more complex, however, in reality this self-consistency is the key to their solution.

We first assume that there exists an effective scalar potential $V_s^{eff}$ that has the same effect as the three potentials combined:
\begin{eqnarray}
\begin{aligned}
\label{eq:fdiraceff}
\left( 1-V_s^{eff}/E-m/E\right)f(x)&=-\frac{g(x)}{x}-g'(x),
\\
 \left( 1+V_s^{eff}/E+m/E\right)g(x)&=-\frac{f(x)}{x}+f'(x).
\end{aligned}
\end{eqnarray}
If such a potential exists, then it must satisfy:
 \begin{equation}
 \label{eq:Vseff}
V_s^{eff}=V_s-V_T\frac{S_2}{S_1} \mbox{ or  }  V_s^{eff}=V_s+V\frac{S_2}{S_0}.
\end{equation}
Inserting the explicit expressions Eqs.\ (\ref{eq:Vs1}) for the potentials in the first identity, we get:
\begin{equation}
\label{eq:Vseff2}
\frac{V_s^{eff}}{E}=-\frac{4C}{3x}~\frac{x(f'g-fg')-x(f^2+g^2)-2fg}{f^2-g^2}.
 \end{equation}
If we now re-insert the Dirac equations as expressed in the effective potential, we get:
\begin{equation}
\label{eq:self}
V_s^{eff}(x)=\frac{4C}{3}[V_s^{eff}(x)+m].
\end{equation}
This is a very beautiful result and confirms the Ansatz that the three potentials can be replaced by a single effective scalar potential. This self-consistency demand means that $V_s^{eff}(x)=0$ as $m\rightarrow0$ and $4C/3\neq 1$. This result was already given in Ref.\ \cite{QuarkDressing} (see Eq.\ (110)). Note that for finite $m$ the mass term $m$ would change into a negative mass term $-9/23  m$.

Since $V_s^{eff}=0$ the quark is free inside the volume $x<x_0$, where $x_0$ is determined by $f(x_0)=-g(x_0)$. Hence, the wave functions are given by:
\begin{equation}
\label{eq:fg}
f(x)=N\sin x;g(x)=N\left(\cos x-\sin x/x\right);
\end{equation}
with normalization:
\begin{equation}
\label{eq:fgnorm}
N^{-2}=x_0-\sin^2 x_0/x_0;x_0=2.04278694\cdots.
\end{equation}	

This solution looks exactly like the idealized model of a particle confined to an infinite well, which is so popular in quantum mechanics
text books. It also agrees with the basic hypothesis of the MIT bag quark model \cite{MIT} and yields the same wave functions with the same domain $[0,x_0]$. There are other analogies with the MIT bag model, however, as stated before the physical context is entirely different, as the MIT bag model describes three quarks in a bag (a nucleon), rather than a single one. Also, the (particle physics) scale of the bag model is entirely different from the Planck scale in the dressing system. Nonetheless it is very interesting that so many properties which were postulated in the phenomenological MIT bag model and now emerge from an exact QFT calculation.

\begin{figure}[!htp]
\begin{center}
\includegraphics[scale=0.45]{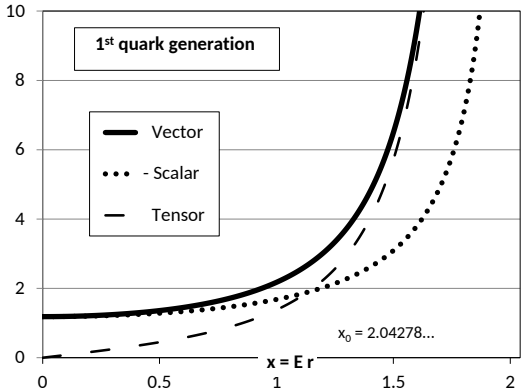}
\caption{Potentials for light quarks}
\label{fig:figpot1}
\end{center}
\end{figure}

\begin{figure}
\includegraphics[scale=0.45]{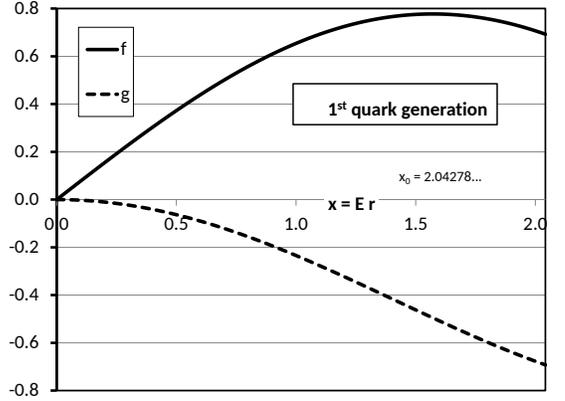}
\caption{Large and small components of the light quark wave function}
\label{fig:figwfs1}
\end{figure}

Now that the wave functions are known we can calculate the original three potentials explicitly. They are shown in Fig.\ \ref{fig:figpot1}.
The negative scalar potential is shown with opposite sign.
The large and small components of the quark Dirac wave function are displayed in Fig.\ \ref{fig:figwfs1}.

III. $\beta=-1;\gamma=1$. Second generation.
\vspace{1.5mm}

\begin{figure}[!ht]
\begin{center}
\includegraphics[scale=0.45]{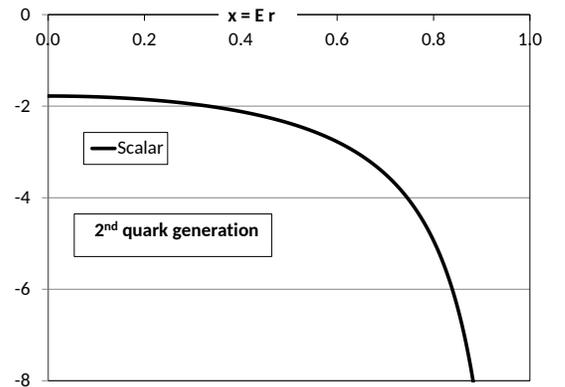}
\caption{2\textsuperscript{nd} generation potential}
\label{fig:potential2}
\end{center}
\end{figure}

\begin{figure}[!ht]
\begin{center}
\includegraphics[scale=0.45]{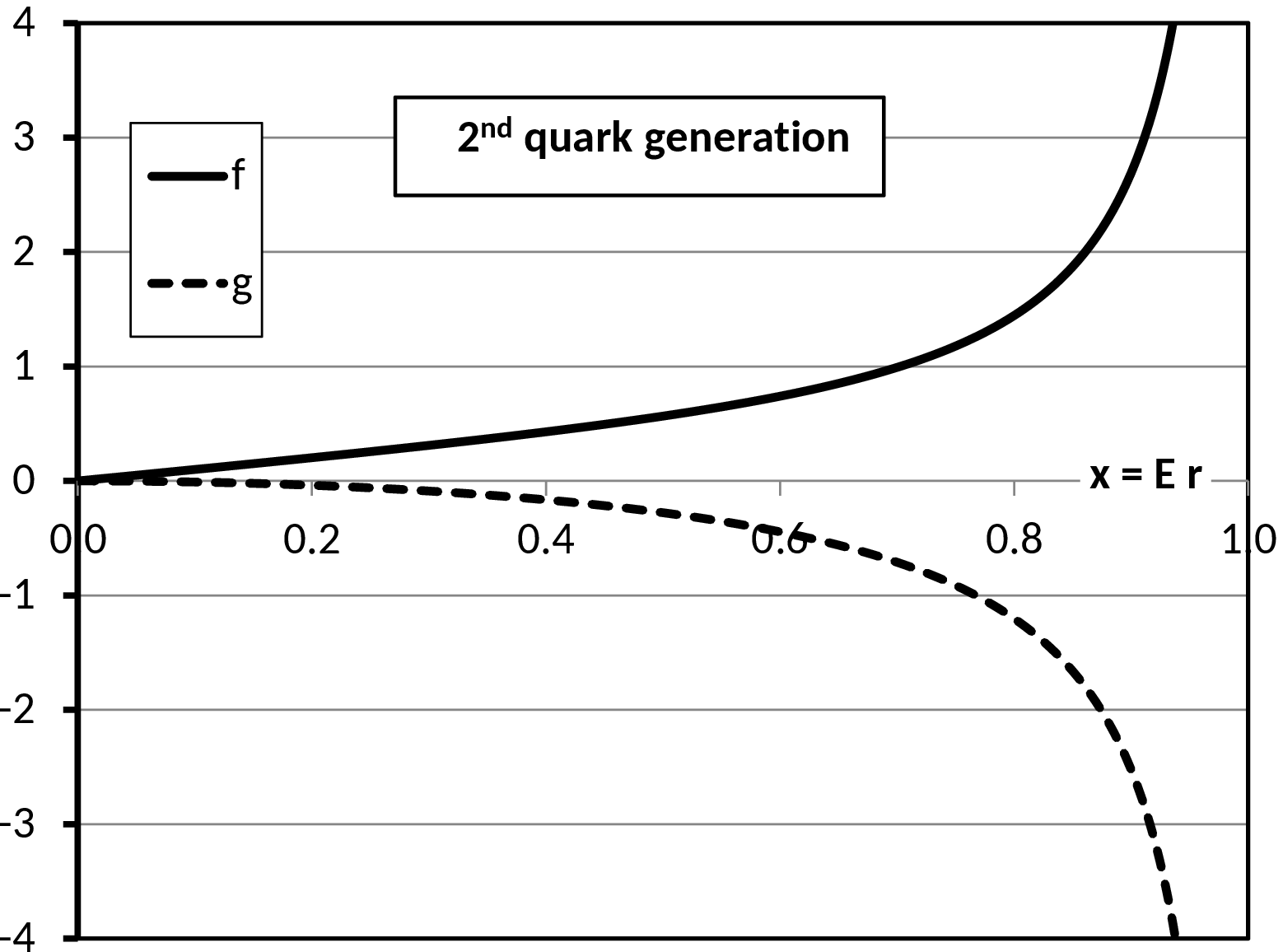}
\caption{2\textsuperscript{nd} generation wave functions}
\label{fig:wfs2}
\end{center}
\end{figure}

\noindent
The exponential function $\hat{E}$ now becomes:
\begin{equation}
\label{eq:Ehat2}
\hat{E}=\frac{S_0+xS_1+S_1+xS_0}{(-x+1)S_+}=\frac{1+x}{1-x}.
\end{equation}		
Hence this solution is singular at $x=1$. In our previous paper \cite{QuarkDressing} we assumed -- incorrectly -- that this singularity
 would be fatal and would prevent the construction of a proper dressing solution. From (\ref{eq:EV}) we find that the vector
 and tensor potential vanish, while the scalar potential is given by:
\begin{equation}
\label{eq:Vs2}
\frac{V_s(x)}{E}=-\frac{2C}{3}\frac{1}{1-x^2}=-\frac{16}{9}\frac{1}{1-x^2}.
\end{equation}
Hence, in this case the potential is not dependent on the wave functions and is already the effective one which can be used in the solution process. This potential is shown in Fig.\ \ref{fig:potential2}.
The wave functions now have the following behaviour near $x=x_0=1$:
\begin{equation}
\label{eq:fg2}
f(x)\sim \frac{1}{(1-x^2)^{8/9}}; g(x)\sim -\frac{1}{(1-x^2)^{8/9}}.
\end{equation}
Just like for the $1^{\text{st}}$ generation we have $f(x)/g(x)\rightarrow -1$ for $x\rightarrow x_0=1$. However, contrary to the first
generation, the ratio of their derivatives approaches -1, rather than +1. The asymptotic behaviour implies that the wave functions
are not normalizable on [0,1]. However, the wave functions can be defined by a limiting procedure for $x_0\rightarrow 1$,
with the normalization constant $N^2$ behaving like $(1-x_0)^{16/9-1}$ when $x_0\rightarrow 1$. For most applications the
functional behaviour is more relevant than the normalization of the wave functions. Hence, all relevant physical properties are still calculable and unaffected by this singular behaviour. However, physically this dressed state can be considered as a two dimensional state with all the quark density concentrated on the surface. The gluon fields remain well-defined over the whole volume $x\leq x_0$, as they are not sensitive to
the wave function normalization. The wave functions are shown for a particular choice of the small $x$-behaviour in Fig.\ \ref{fig:wfs2}.

IV. $\beta=1;\gamma=-1$. Third generation.

\noindent
The function $\hat{E}$ now equals:
\begin{equation}
\label{eq:Ehat3}
\hat{E}=\frac{1-x}{1+x}\frac{S_-}{S_+}.
\end{equation}	
This looks like a mixture of the first and second generation results. The resulting potentials also look like mixtures of the two previous cases:	\begin{eqnarray}
\begin{aligned}
\label{eq:Vs3}
\frac{V(x)}{E}&=\frac{4C}{3x}\frac{f^2+g^2}{(f^2-g^2)^2}\left[x(f^2+g^2)+2fg\right],
\\
\frac{V_s(x)}{E}&=-\frac{4C}{3}\frac{f'g-fg'}{f^2-g^2}+\frac{2C}{3}\frac{1}{1-x^2},
\\
\frac{V_T(x)}{E}&=-\frac{8C}{3x}\frac{fg}{(f^2-g^2)^2}\left[x(f^2+g^2)+2fg\right].
\end{aligned}
\end{eqnarray}

In comparison to Eq.\ (\ref{eq:Vs1}) there is an extra term in the scalar potential, which -– apart from the sign -- is identical to the scalar potential for the $2^{\text{nd}}$ generation.
We can now carry out the same reduction as we did for the first generation by introducing an effective scalar potential, as the extra scalar term does not depend on the wave functions. Hence, we get the same renormalization formula Eq.\ (\ref{eq:self}) as before, with the mass term $m$ replaced by the extra scalar potential. The scalar potential equivalent to these three potentials (excluding the extra scalar term) then becomes:
\begin{eqnarray}
\begin{aligned}
\label{eq:self3}
V_s^{equiv}(x)&=\frac{4C}{3}\left[V_s^{equiv}(x)+\frac{2C}{3}\frac{1}{1-x^2}\right]\Rightarrow
\\
V_s^{equiv}(x)&=\frac{4C/3}{1-4C/3} \frac{2C}{3} \frac{1}{1-x^2}.
\end{aligned}
\end{eqnarray}
Adding the extra scalar potential we get:
\begin{equation}
\label{eq:Veff3}
V_s^{eff}(x)=\frac{2C/3}{1-4C/3}\frac{1}{1-x^2}=- \frac{16}{23}\frac{1}{1-x^2},
\end{equation}
where we kept the label \emph{eff} for this sum.
Although the extra scalar potential in Eq.\ (\ref{eq:Vs3}) had a positive sign, the renormalization procedure has inverted it. Hence, the effective scalar potential does not only have the same functional form as the potential for the second generation, but it also has the same sign. Only its magnitude has been reduced by the factor 9/23 mentioned before.

\begin{figure}[!ht]
\begin{center}
\includegraphics[scale=0.45]{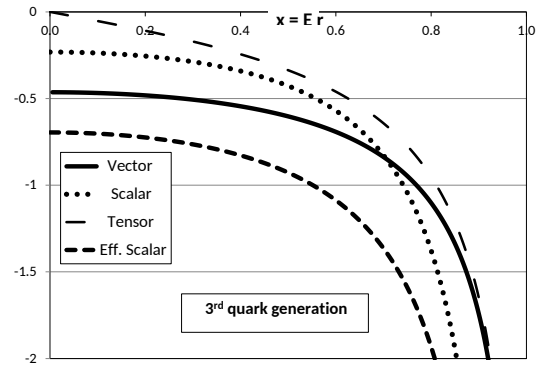}
\caption{3\textsuperscript{rd} generation potentials}
\label{fig:potentials3}
\end{center}
\end{figure}

\begin{figure}[!ht]
\begin{center}
\includegraphics[scale=0.45]{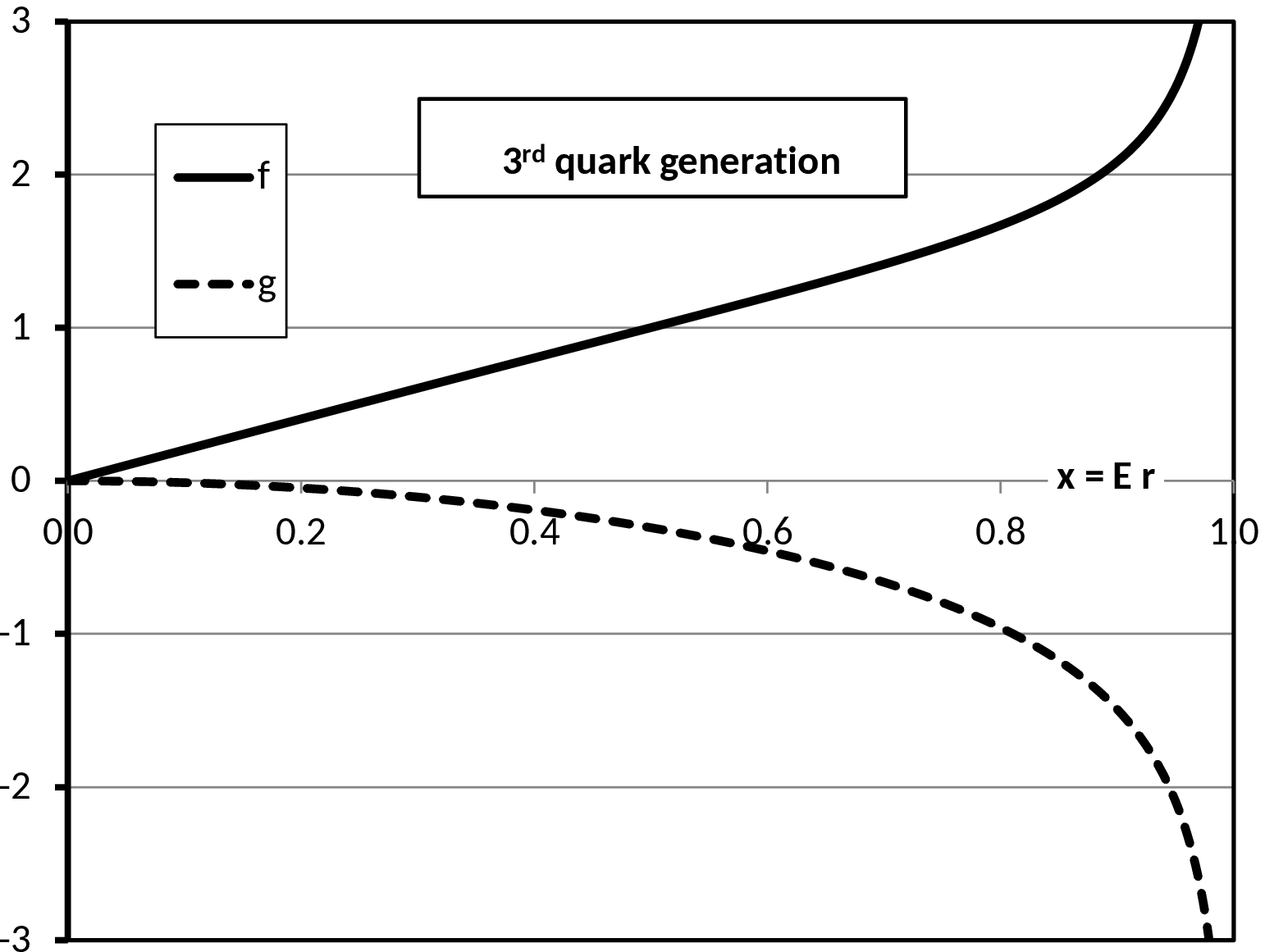}
\caption{3\textsuperscript{rd} generation wave functions}
\label{fig:wfs3}
\end{center}
\end{figure}

The wave functions for the third generation have the same asymptotic behavior, Eq.\ (\ref{eq:fg2}), as they do for the second generation, except that  the coefficient is reduced from 8/9 to 8/23. So the wave functions are now normalizable without the need for limiting procedures, although they still approach infinity for $x\rightarrow 1$. In Fig.\ \ref{fig:potentials3} we show the original and the effective scalar potential.
The wave functions can be normalized in the standard way and are shown in Fig.\ \ref{fig:wfs3}.

The three dressing solutions have very distinct characteristics. The quarks belonging to the first generation are confined by an infinite well, but free inside;  quarks belonging to the second generation live on the surface, so that the normalization must be carried out via a limiting procedure; while for the third generation the bound state energy cancels the bare energy $E$, so that the total energy vanishes, as we will see in the next section. Each of the three solutions therefore represents some extreme limiting situation which gives it a unique character. At the same time all four solutions are related by symmetries.

Since these symmetries accentuate the elegance and structural nature of the solutions we want to close this section with a simple example of these properties. For all three generations the effective potential is of scalar type and can even be represented by the same functional form:
\begin{equation}
\label{eq:Vseff234}
V_s^{eff}(x)=-\frac{a^{(n)}}{1-x^2};
a^{(n)}=\left(0, 0, \frac{16}{9},\frac{16}{23}\right),
\end{equation}
where we have included the bare case, as well. The pair (2,3) satisfies the symmetry relation:
\begin{equation}
\label{eq:symmetry}
a^{(3,2)}=-\frac{a^{(2,3)}}{1-2 a^{(2,3)}}.
\end{equation}
For the pair $n=(0,1)$ this relationship is also (trivially) satisfied. The simplicity and elegancy of these results, emerging from the originally enormously complicated operator equations, is testimony to the strength of the exact approach which is possible in the dressing theory.

\section{Total QFT energy of dressed quarks}
\label{sec:TotalEnergy}
One of the remarkable aspects of the dressing theory is the nearly complete separation of the dynamical field calculations and the energy calculations. In this respect QFT is very different from non-relativistic Schrodinger calculations, where the determination of the energy eigenvalue is an inherent part of the dynamical calculation, even characterizing the different (eigen-) solutions. In QFT the total energy can be determined \emph{after} the dynamical calculations have been completed by calculating the expectation value of the relevant operators using the previously determined fields. The total QFT energy equals:
\begin{equation}
\label{eq:Etot1}
E^{\alpha (n)}_{tot}=\int d^3r\left\langle \alpha n\right|\theta^{00}\left| \alpha n\right\rangle ,
\end{equation}
where $\theta^{00}$ is the energy-component of the symmetrized energy-momentum tensor $\theta^{\mu\nu}$ \cite{Peskin}:
\begin{equation}
\label{eq:Theta}
\theta^{\mu\nu}=-\mathscr{L}g^{\mu\nu}+\frac{\textrm{i}}{2}\bar{\psi}
(\gamma^\nu D^\mu+\gamma^\mu D^\nu)\psi-F_a^{\mu\rho}F_{a,\rho}^\nu,
 \end{equation}
and the dressed quark state is given by the state vector:
\begin{equation}
\label{eq:state_vector}
\left| \alpha n\right\rangle=\left| b_{\alpha n}^\dag\right|0\rangle; n=0,\cdots 3.
\end{equation}
Here $n$ labels the generation quantum number, i.e. it specifies which field solutions $(\psi,A^\mu_a)$ and profile functions should be used in the calculation of the expectation values. As usual $\alpha$ labels the spin and colour of the dressed state. Since the expectation value does not depend on these quantum numbers, $\alpha$ will be omitted in the following.


The energy expression can be simplified by imposing the Dirac equation (we temporarily abandon the dimensionless representation with $x$):
\begin{eqnarray}
\label{eq:Etot2}
\begin{aligned}
& E^{(n)}_{tot}=E+\frac{16}{3}\frac{g_s}{2}
\int d^3r\left\langle n\right|\bar{\psi}\gamma_0\psi A^0\left| n\right\rangle+E_{gluon}^{(n)},
\\
& E_{gluon}^{(n)}=\frac{16}{3}
\int d^3r\left\langle n\right|\frac{1}{4}F^{\mu\nu}F_{\mu\nu}- F^{0i}F_{0i}  \left|  n\right\rangle,
\end{aligned}
\end{eqnarray}
where we carried out the summation over the color index $a$. The first term $E$ is simply the energy of a bare quark with time dependence $\exp(-iEt)$ or an anti-quark with time dependence $\exp(iEt)$. The other terms are due to the quark-gluon interaction and the presence of the gluon fields and are expected to give an overall negative contribution to the total (binding) energy.

We rewrite this result in terms of the reduced profile functions, so that we can introduce the solutions. Since the quadratic gluon term vanishes in the limit $\alpha_s\rightarrow0$, we ignore it from now on. For a discussion of this $\mathcal{O}(\alpha_s)$ term we refer to \cite{QuarkDressing}. We now get the following result for the total energy:
\begin{equation}
\label{eq:Etot3}
E^{(n)}_{tot}=E\left[1-\frac{16}{9}+\frac{16}{9}\frac{\beta}{|\beta|}\frac{2}{\alpha_s}\int^{x_0}_0 dx x \hat{S}_0\right].
\end{equation}
In the transition to reduced functions the original source functions are replaced by hatted ones. In analogy to Eq.\ (\ref{eq:S+S-}) the hatted source function $\hat{S}_0$ is defined by:
 \begin{equation}
\label{eq:S0hat}
\hat{S}_0=\frac{\hat{S}_++\hat{S}_-}{2}=S_0\frac{\hat{E}+\hat{E}^{-1}}{2}+S_1\frac{\hat{E}-\hat{E}^{-1}}{2}.
\end{equation}
Now we discuss the results for the four solutions defined in Sec.\ \ref{sec:wave functions}.

For the bare solution we have $\beta=\gamma=1$, so that $\hat{E}=\hat{E}^{-1}=1$ and $\hat{S}_0=S_0$. We then get:
\begin{equation}
\label{eq:Etot4}
E^{(0)}_{tot}=E\left[1-\frac{16}{9}+\frac{16}{9}\right]=E.
\end{equation}
As expected the bare solution only features the bare energy term $E$.

The first dressing solution is given by $\beta=\gamma=-1$ and $\hat{E}=S_-/S_+$. Again we have $\hat{S}_0=S_0$, so that:
\begin{equation}
\label{eq:Etot5}
E^{(1)}_{tot}=E\left[1-\frac{16}{9}-\frac{16}{9}\right]=-\frac{23}{9} E.
\end{equation}

The second generation solution is given $\beta=-1;\gamma=1$ and $\hat{E}=(1+x)/(1-x)$, so that:
\begin{eqnarray}
\begin{aligned}
\label{eq:S0hat2nd}
\hat{S}_0&=\frac{S_0(1+x^2)+2xS_1}{1-x^2}.
\\
&\equiv \frac{1-x}{1+x}S_0 +\frac{2x}{1-x^2}S_+.
\end{aligned}
\end{eqnarray}
After careful analysis we find that $f+g\sim (1-x)^{8/9}$ near $x=1$, so that the integral over $S_+/(1-x^2)$ converges as $S_+\sim (f+g)^2$. Since the normalization constant $N(x_0)$ goes to zero for $x_0\rightarrow 1$, the second term does not contribute to the energy. The same is true for the first term, as the integral over $S_0(1-x)\sim (1-x)^{-7/9}$ converges as well. Hence, there is no contribution of the last term in Eq.\ (\ref{eq:Etot3}) and we obtain:
\begin{equation}
\label{eq:Etot6}
E^{(2)}_{tot}=E\left[1-\frac{16}{9}\right]=-\frac{7}{9} E.
\end{equation}

The third generation is given by $\beta=1;\gamma=-1$ and $\hat{E}=[(1-x)S_-]/[(1+x)S_+]$. Just like for the $2^{\text{nd}}$ generation we find:
\begin{equation}
\label{eq:S0hat3rd}
\hat{S}_0=\frac{S_0(1+x^2)+2xS_1}{1-x^2}.
\end{equation}
We now follow a slightly different route to calculate the total energy. This reasoning is also valid for the second generation, so we include this case as well. Suitably combining the Dirac equations for $f$ and $g$  in Eq.\ (\ref{eq:fdiraceff}), and inserting the general formula Eq.\ (\ref{eq:Vseff234}) for the effective scalar potential, one can rewrite $\hat{S}_0$ as follows:
\begin{eqnarray}
\begin{aligned}
\label{eq:S0hat3rdx}
& x\hat{S}_0=-xS_0+2x\frac{S_0+xS_1}{1-x^2}=
\\
& =-xS_0-\frac{1}{a^{(n)}}\left[(xS_1)'+x(xS_0)')\right].
\end{aligned}
\end{eqnarray}
Substituting this in Eq.\ (\ref{eq:Etot3}) we get:
\begin{eqnarray}
\begin{aligned}
\label{eq:E23}
& \frac{E^{(2,3)}_{tot}}{E}=\left[1-\frac{16}{9}-\frac{16}{9}\frac{\beta}{|\beta|}\right]\int_0^{x_0}dx(f^2+g^2)
\\
& -\frac{1}{a^{(2,3)}}\frac{16}{9}\frac{\beta}{|\beta|}\int_0^{x_0}dx \left[(2fg)'+x(f^2+g^2)'\right].
\end{aligned}
\end{eqnarray}
Inserting the actual values of $\beta$ and $a^{(n)}$ we find that both cases are proportional to the same complete integral:
\begin{eqnarray}
\begin{aligned}
\label{eq:E23concise}
& \frac{E^{(2,3)}_{tot}}{E}=(1,-\frac{23}{9})\int_0^{x_0}\left[x(f^2+g^2)+2fg\right]'=
\\
& =(1,-\frac{23}{9})\lim_{x \rightarrow x_0}\left[x(f+g)^2+2fg(1-x)\right].
\end{aligned}
\end{eqnarray}
For the $3^{\text{rd}}$ generation the wave function is normalizable and the limit yields zero, so that the total energy vanishes as the interaction energy exactly cancels the bare energy $E$. For the $2^{\text{nd}}$ generation the limit is finite and we get the result already given in Eq.\ (\ref{eq:Etot6}), namely $-7/9 E$. Eq.\ (\ref{eq:E23concise}) again illustrates the striking symmetry between the $2^{\text{nd}}$ and $3^{\text{rd}}$ generation, and the remarkable elegance of these results.

The results can be summarized as follows:
\begin{eqnarray}
\label{eq:Esummary}
E^{(n)}_{tot}= -\gamma^{(n)}E;\; \gamma^{(n)}=\left(\frac{23}{9},\frac{7}{9},0\right).
\end{eqnarray}
We note that the differences between the total (binding) energy and the bare energy $E$, namely $(32/9,16/9,1)E$, relate directly to the SU(3) structure constants, re-emphasizing the structural nature of the dressing solutions.
\section{Mass calculation of dressed quarks}
\label{sec:Mass}
In order to determine the masses of the dressed quarks we have to address two fundamental problems. First the mass of dressed quarks cannot be equated to the total energy $E^{(n)}_{tot}$, as the latter is negative (or zero for $n=3$). Second, the QFT dressing theory is scale invariant, so the energy scale $E$ remains to be determined.

This issue of scale seems to lead to conflicting demands. On the one hand we expect a dressed quark (if it is to be finite at all) to be very small, possibly even as small as the Planck length, which is perhaps the only fundamental length scale. On the other hand, the empirical mass of a light quark is of the order of a few MeV. But $E\times r_0=x_0$ is of the order of unity, so if $r_0$ is of the order of Planck lengths then $E$ is of the order of Planck energies. Since the total energy $E^{(n)}_{tot}$ is expressed in units of $E$, we would expect that the quark mass is of the same order, in flagrant contradiction with observation. This potential conflict was a continuous concern in our research project and might well have discouraged others from  contemplating such a spatial model at all, despite its physical appeal.

There is a remarkable escape from this puzzle which exploits the negativity of the total (binding) energy.
First we assume that this negativity is a general property of dressed states and that the case $n=3$, for which the energy was zero, can be treated as a limiting case with $E^{(3)}_{tot}$ approaching zero from below, making the tacit assumption that this energy will also turn negative as soon as more contributions (e.g. of other interactions) are included.

The negative energy of the dressed state can be lowered without bound by increasing $E$, or equivalently by reducing $r_0$. However, at some point during this collapse, $E$ becomes of the order of the Planck energy and the effects of general relativity kick in. While normally GR would accelerate such a collapse, in the case of negative energy it will resist it and the total energy will stabilize at a certain value of $E$.

To quantify this effect the spatial integral energy integral is modified by introducing the extra measure $\sqrt{|g_{\mu\nu}|/|g_{00}|}$, where $g_{\mu\nu}$ is based on a black hole metric (naturally with inverted sign for the energy/mass). Using the isotropic metric (see \cite{Weinberg}, p. 181) and keeping only terms to first order in $G$, we find:
\begin{eqnarray}
\label{eq:EtotGR}
\begin{aligned}
&E^{(n)}_{tot}\rightarrow\int d^3r \left(1+3\frac{GE^{(n)}_{tot}}{r}\right)\left\langle \alpha n\right|\theta^{00}\left| \alpha n\right\rangle
\\
&=-\gamma^{(n)}E\left( 1-3 \gamma^{(n)} GE^2\left\langle\frac{1}{x}\right\rangle\right),
\end{aligned}
\end{eqnarray}
which is identical to the expression given in \cite{QuarkDressing}. The coefficient 3 was obtained using the isotropic metric, but could be different for different metrics, as each metric uses different definitions of the space-time variables (for a survey see \cite{Weinberg}). The non-uniqueness of black hole metrics is a consequence of the fact that they are designed to match the long-range Newtonian behavior, but can differ in the interior region, which is the region of interest to us.

Various modifications and/or improvements are possible on this procedure. For example, one can replace the total energy $E^{(n)}_{tot}$ in the correction factor by an energy integral up to $r$ instead of $r_0$, although this requires a unique definition of the local energy density. One could also consider higher-order terms in $G$, however, without a more complete unification of QFT and GR their effects might be misleading. These modifications would in principle be testable as they affect the mass calculations. However, until we have included the important electro-weak and Higgs interactions such tests may not be very conclusive. What is important though, is to appreciate the vital role of the negativity of the QFT energy. We do not expect this property to change once other interactions are introduced, although the third generation of quarks could have a special status, as it currently features zero binding energy.

We now minimize the modified energy expression, Eq.\ (\ref{eq:EtotGR}), with respect to $E$, thereby fixing its value:
\begin{equation}
\label{eq:E_Planck}
E\rightarrow E^{(n)}=\frac{1}{3}\frac{<1/x>^{-1/2}}{[\gamma^{(n)}G]^{1/2}}=\frac{1}{3}\left[\frac{\delta^{(n)} x_0^{(n)}}{\gamma^{(n)}G}\right]^{1/2},
\end{equation}
where according to \cite{QuarkDressing} $\delta^{(1)} \approx0.6019$. The corrected total energy for the $\text{n}^{\text{th}}$ dressed state now becomes:
\begin{equation}
\label{eq:EtotGRfinal}
E^{(n)}_{tot}\rightarrow-\frac{2}{3}\gamma^{(n)}E^{(n)}.
\end{equation}

While we have fixed $E$, we have not yet found a positive energy expression which can be identified with the mass. In non-relativistic physics the overall mass of a system is obtained by subtracting the binding energy from the total sum of the individual masses (see Eq.\ (2-11) in \cite{Bohr} for the nuclear case). For a dressed quark such a basis is absent and we have to find a different origin of the ground or vacuum state with positive energy.

A vacuum term of quantum field theoretic origin, such as a chiral condensate or the vacuum term proposed in the MIT bag model \cite{MIT}, would not work in the current case. The reason is that in order to represent the mass of the dressed system the resulting total energy would have to be positive, thereby invalidating the earlier arguments. Hence, we are looking for an inert background term which does not change the dynamics of the QFT system and does not participate in the energy minimization, so that it preserves the result Eq.\ (\ref{eq:E_Planck}). The energy density of such a vacuum term would have to be enormous to compensate for the (negative) energy density of order $\mathcal{O}(G^{-2})$. To match this density requires circumstances only believed to have existed in the early universe. This has led us to the following proposal.

The simplest non-trivial vacuum universe with a positive energy content is an empty universe characterized by a positive cosmological constant $\Lambda$. Such de Sitter universes \cite{Sitter} have been studied extensively (see \cite{Weinberg} p. 615). However, when expressed in terms of a conformal metric:
\begin{equation}
\label{eq:guv}
g_{\mu\nu}=g(t)\eta_{\mu\nu},
\end{equation}
they become especially useful for our purposes. The time $t$ defined in this way is known as the conformal time.
Solving the Einstein equations for this metric one finds (\cite{GRcosmology}, \cite{Grebenbook}):
\begin{equation}
\label{eq:g}
g(t)=(t_s/t)^2;\quad t_s=\left(\frac{3}{8\pi G\epsilon}\right)^{1/2}=\left(\frac{3}{\Lambda}\right)^{1/2},
\end{equation}
where the vacuum energy density is given by the expression $\epsilon=\Lambda/8\pi G$.

If one uses this as a cosmological model for the whole universe then one finds that the vacuum energy density should equal $\epsilon=3.97\times 10^{-47}~\textrm{GeV}^4$ in order to fit the supernovae data \cite{Riess1}, \cite{Riess2}. This is equivalent to a cosmological constant $\Lambda= 6.69\times 10^{-84}~\textrm{GeV}^2$. One important property of the conformal parametrization is that the geodesics are unchanged from the Minkowski ones, except that they are confined to $t>0$ or $t<0$, where $t=0$ characterizes the birth of the universe and the two branches reflect the time symmetry of the Einstein equations. Since the Minkowski metric and its geodesics are used throughout QFT, we feel that this minimal conformal metric is a suitable background metric, and may well qualify as the inert vacuum background we are trying to emulate. This vacuum universe must be contained inside the spherical boundary of the dressed system, so that it should terminate its expansion at a time (which we will call $t_c$), when its extent matches the volume of the dressed system.

In order to calculate the energy content of this vacuum universe we must integrate the energy density over the volume of the dressed system, taking account of the conformal metric in the spatial integral. The metric can be absorbed in the effective energy density, leading to:
\begin{equation}
\label{eq:vacuum}
\epsilon\rightarrow\epsilon_{eff}(t)=\sqrt{g^3(t)}~\epsilon=(t_s/t)^3~\epsilon.
\end{equation}
At the time of termination the energy density then reads $(t_s/t_c)^3~\epsilon$. The combined energy of the dressed system -- which is identified with the quark mass $m_q$ -- now equals:
\begin{eqnarray}
\label{eq:tc}
m_q=\frac{4\pi}{3}\left(\frac{x_0^{(n)}}{E^{(n)}}\right)^3\epsilon \left(\frac{t_s}{t_c}\right)^3+  E^{(n)}_{tot}.
\end{eqnarray}
Since there are two unknowns in this equation, namely $t_c$ and $m_q$, we cannot yet solve these equations.

We now make the crucial assumption that the time and energy needed to create this state are related by the Heisenberg uncertainty relation, i.e.
\begin{equation}
\label{eq:mqtc}
m_q=\frac{1}{2}\frac{1}{t_c},
\end{equation}
where as usual we have set $\hbar=1$. Physically we can motivate this assumption by noting that the external environment can only support an energy bubble of magnitude $m_q$ for a period $t_c$. Once the system has been formed it becomes stable and the supplied energy will be extracted permanently from the environment, for example through a scattering process. The Heisenberg uncertainty principle is often used to describe the decay of nuclear states and their decay width, however, their use for stable states is less common. We can also consider Eq.\ (\ref{eq:mqtc}) as a testable working hypothesis, as it leads to concrete predictions for the quark masses.

Imposing Eq.\ (\ref{eq:mqtc}) on Eq.\ (\ref{eq:tc}) converts it into a self-consistent equation for $t_c$ or for the mass of the dressed quark. Since the mass of the quark is much smaller than the two contributions in the RHS, there must be a nearly complete cancellation between the positive vacuum energy and the negative QCD energy. This yields the identity:
\begin{equation}
\label{eq:tcreation}
t_c=\frac{9}{2}\left(\frac{G}{6\pi \epsilon}\right)^{1/6}\left(\frac{x_0^{(n)}\gamma^{(n)}}{\delta^{(n)2}}\right)^{1/3},
\end{equation}
so that the quark mass $m_q^{(n)}$ is given by:
\begin{equation}
\label{eq:mq}
m_q^{(n)}=\frac{1}{9}\left(\frac{6\pi \epsilon}{G}\right)^{1/6}\left(\frac{\delta^{(n)2}}{x_0^{(n)}\gamma^{(n)}}\right)^{1/3}.
\end{equation}
The quark mass is now expressed in terms of the fundamental physical constants $G$ and $\epsilon$ and its order of magnitude is given by
\begin{equation}
\label{eq:scale}
(6\pi \epsilon/G)^{1/6}= (3 \Lambda/4 G^2)^{1/6}= 69.38 \text{ MeV}.
\end{equation}
Hence, the dressing formulation, in combination with the cosmological model, has provided a long-sought connection between cosmological parameters and particle physics. This is exactly the result we were hoping for initially, though its realization looked a dim prospect for a long time. To reach this stage required a number of large conceptual steps, which however now find strong support in this result.

If we now apply Eq.\ (\ref{eq:tcreation}) and Eq.\ (\ref{eq:mqtc}) to the light quarks ($x_0^{(1)}=2.04279$; $\gamma^{(1)}=23/9$; $\delta^{(1)}=0.6019$) we find that $m_q= 3.17$ MeV; a result first presented in Ref.\ \cite{QuarkDressing}. This result is even more amazing than the order of magnitude estimate given above, since it lies right in the empirical range found in lattice gauge calculations, namely $3.8 \pm 0.8$ MeV \cite{Lattice Quark masses}. However, we view this further success as a bit fortuitous, as we would not expect such an accurate answer while having ignored the electroweak and Higgs interactions.

Applying the same formulae to the second generation ($x_0^{(2)}=1; \gamma^{(2)}=7/9; \delta^{(2)}=1$) yields a mass of 8.38 MeV. Although this is considerably larger than the value obtained for the light quarks (by a factor 2.6), the result falls far short of the observed values.

For the third generation the limiting procedure $\gamma^{(3)}\downarrow 0$ leads to an infinite quark mass. This may not be a bad initial result in view of the enormous quark masses for this generation. Clearly, further binding is necessary to make $\gamma^{(3)}$ positive and produce a finite mass. Since the masses for the higher generations depend strongly on charge, accurate results can only be expected after the missing electroweak interactions are included.

In order to introduce electroweak and Higgs interactions in the dressing theory, one needs to formulate a basic Lagrangian without reference to the generation quantum number. Because of the complexity of the electroweak interactions and the role of spontaneous symmetry breaking in the SM formulation, the reduction of the SM electroweak Lagrangian to such a basic form is not so straightforward as it was for QCD. However, we have experimented with a simple Higgs interaction which does not refer to the generation quantum number. Our tentative results suggest that the Higgs interactions are suppressed for the light quark case because of certain cancellations in the first generation solution. If substantiated, this could be an explanation for the observation that the Higgs mainly interacts with the higher generations.

Apart from improvements in the dressing theory there may also be more conventional standard QFT scattering diagrams (external to the dressed state) which contribute to the effective quark mass, so that the quark masses we calculate cannot directly be compared to the observed ones. As stated in the introduction, this paper does not address the integration of the theory in the SM framework, where questions of such additional contributions might arise.

\section{Summary and future developments}
\label{sec:conclusions}
In this paper we presented a theory of the dressing of bare quarks based on the solution of the quantized (QCD) field equations. Using this approach we were able to prove that the gluon fields generated by the bare quark lead to the absolute confinement of the quark and gluon fields within a small spherical volume of Planck length dimensions. With its ability to explain the existence of the three generations and its potential to predict the quark masses in terms of a few fundamental constants of nature, this theory could become an important foundation block for particle physics.

The success of this theory was crucially dependent on a number of theoretical discoveries and could only be completed after certain consequences of general relativity were accounted for and a vacuum background term -- inspired by cosmology -- was included.

The first important discovery was that the quantized operator equations, expressed in terms of creation and annihilation operators, admitted an exact operator solution. This solution defined its own domain of applicability by limiting the allowed state vectors to single-particle quark states, thereby identifying the dressing problem as a unique problem in QFT that requires a unique methodology. This result makes it more plausible why scattering methods in QFT have been so dominant from the start and why the dressing application has taken so long to be discovered, although this could also be due to the apparent conflict between small quark masses and their non-zero, but very small size.

The knowledge of the exact operator solution enabled the reduction of the quantized field equations to c-number equations. This paved the way for further discoveries. The reduced coupled field equations allowed three structural (eigen)solutions, which even survived in the limit $\alpha_s\rightarrow0$. Hence, the likely reason behind the existence of three generations of quarks was identified. Within the dressed system the gluon interactions -- though originally of vector type -- translated in scalar, vector and tensor interactions. Subsequently these potentials could be reduced to a single effective scalar potential, which gave the formal theory a strong physical appeal. For example, the solutions displayed a strong similarity with the phenomenological MIT three quark bag.

The collapse of the dressed system with negative total energy could be prevented by including GR in the description, leading to a size of the order of Planck lengths. However, this constrained the possible solutions to the problem of the construction of a vacuum term that -- together with the negative binding energy -- could be the basis for mass calculations. Such a term would have to compensate for the enormous (negative) energy density of the dressed system. The solution was to define a vacuum of cosmological signature. After its introduction the quark masses became calculable and were found to lie in the MeV region, despite being a result of the cancellation of terms of the order of the Planck energy. This demonstrated for the first time the important role of cosmological parameters in particle physics.

The dressing theory looks to open a new domain of applications of QFT which brings with it techniques that are more common in non-relativistic physics, such as the concept of discrete eigensolutions, normalized states, binding potentials and bound-state wave functions. Aspects both new to QFT and non-relativistic quantum mechanics are the exact quantization techniques, the strong reliance on self-consistency and the dependency on a successful extension into general relativity and cosmology. Many elements are still missing, such as the inclusion of electroweak and Higgs interactions, so there are ample opportunities for extensions and improvements whose predictions can be tested against the existing knowledge of elementary particles.

The powerful techniques and methodologies developed in the dressing theory, in particular the exact operator techniques, could also benefit the standard scattering applications in QFT and the SM. To use them in the case of boson fields, one should expand these fields in bilinear (bare) quark-anti-quark operators interspersed if appropriate with the infinite operator $\Lambda_\infty$, just like we did for the dressing case. Naturally, in the scattering case one should employ a continuous spectrum of plane waves (with the momentum shared equally between quark and anti-quark), rather than unknown discrete bound-state wave functions. In such an expansion most physical properties of the fields are automatically built in. By quantizing boson fields this way we were able to avoid the cosmological constant problem \cite{CC problem}, maybe the most serious problem in the conventional quantization approach in QFT.

It also appears possible to extend this methodology to the lepton sector and thus explain the existence of three lepton generations. We hope to discuss this extension in a further publication. One of the important questions which need answering in this connection is why the neutrino masses are so tiny. Like the large quark masses for the higher generations, the explanation of these small masses may well require additional conceptual steps and lead to new theoretical surprises. However, the current theory has shown that it is no longer justified to consider these masses as empirical parameters, but rather as parameters which deserve a theoretical explanation.

\begin{acknowledgments}
The author acknowledges a discussion with Prof. Vincent Icke which contributed to the physical understanding of the energy results. He also acknowledges comments from Prof. Piet Mulders.
\end{acknowledgments}

\end{document}